\documentclass{article}
\usepackage{cancel,graphicx,float}
\usepackage{amsmath}
\usepackage{amsmath,amssymb,latexsym}

\def\la{\langle}
\def\ra{\rangle}
\def\ve{\vert}
\def\Occ{\operatorname{Occ}}

\usepackage{tikz}
\usepackage{booktabs}

\usetikzlibrary{calc,fit,shadows,arrows,positioning}
\pgfdeclarelayer{background}
\pgfdeclarelayer{foreground}
\pgfsetlayers{background,main,foreground}
\definecolor{lightblue}{rgb}{0.96,0.96,1.0}
\definecolor{darkblue}{rgb}{0,0,0.4}

\newsavebox{\dataTableContent} 
\newenvironment{dataTable}[1] 
{%
\begin{lrbox}{\dataTableContent}%
\begin{tabular}{#1}}%
{%
\end{tabular}
\end{lrbox}
\begin{tikzpicture}
\node [inner xsep=0pt] (tbl){\usebox{\dataTableContent}};
\begin{pgfonlayer}{background}
\draw[rounded corners=1pt,top color=lightblue!15,bottom color=darkblue!15,draw=black]
(tbl.north east) rectangle (tbl.south west);
\draw[rounded corners=1pt,top color=darkblue!70,bottom color=darkblue,draw=black]%
($(tbl.north west)$) rectangle ($(tbl.north east)-(0,1.5\baselineskip)$);
\draw[rounded corners=0.25pt,fill=darkblue!80,draw=black]%
(tbl.south west) rectangle ($(tbl.south east)+(0,0.05)$);
\end{pgfonlayer}
\end{tikzpicture}}

\usepackage{authblk}

\title{Ubiquitous nucleosome unwrapping in the yeast genome}
\author[1]{R\u{a}zvan V. Chereji\thanks{rchereji@physics.rutgers.edu}}
\author[1,2]{Alexandre V. Morozov\thanks{morozov@physics.rutgers.edu}}
\affil[1]{Department of Physics and Astronomy, Rutgers University, Piscataway, New Jersey 08854-8019, USA}
\affil[2]{BioMaPS Institute for Quantitative Biology, Rutgers University, Piscataway, New Jersey 08854-8019, USA}

\date{May 17, 2013}

\begin{document}
\maketitle

\begin{abstract}
Nucleosome core particle is a dynamic structure -- DNA may transiently peel off
the histone octamer surface due to thermal fluctuations or the action of chromatin
remodeling enzymes. Partial DNA unwrapping enables easier access of DNA-binding
factors to their target sites and thus
may provide a dominant pathway for effecting rapid and robust access to DNA packaged into chromatin.
Indeed, a recent high-resolution map of distances between neighboring nucleosome dyads in \textit{S.cerevisiae} shows that at least 38.7\% of all nucleosomes are partially
unwrapped. The extent of unwrapping follows a stereotypical pattern in the vicinity of
genes, reminiscent of the canonical pattern of nucleosome occupancy in which nucleosomes
are depleted over promoters and well-positioned over coding regions. To explain these
observations, we developed a
biophysical model which employs a 10-11 base pair (bp) periodic nucleosome energy profile. The profile, based on the pattern of histone-DNA contacts in nucleosome crystal structures and the idea of linker length discretization, accounts for both nucleosome unwrapping and
higher-order chromatin structure. Our model reproduces the observed genome-wide distribution of inter-dyad distances, and accounts for patterns of nucleosome occupancy and unwrapping around coding regions.
At the same time, our approach explains \textit{in vitro} measurements of accessibility
of nucleosome-covered binding sites, and of nucleosome-induced cooperativity between DNA-binding factors. We are able to rule out several alternative scenarios of nucleosome unwrapping as inconsistent with the genomic data.
\end{abstract}

Eukaryotic genomes are organized into arrays of nucleosomes~\cite{vanholde:1989}. Each nucleosome consists of a stretch of genomic DNA wrapped around a histone octamer core~\cite{Luger1997}. The resulting complex of DNA with histones and other regulatory and structural proteins is called chromatin~\cite{vanholde:1989,felsenfeld:2003}. Arrays of nucleosomes form 10-nm fibers which resemble beads on a string and fold into higher-order structures~\cite{Olins1974,Kornberg1974,Woodcock1976}.
Depending on the organism and cell type, 75-90\% of genomic DNA is packaged into nucleosomes~\cite{vanholde:1989}.
Since nucleosomal DNA is tightly wrapped around the histone octamer, its accessibility to various DNA-binding proteins such as repair enzymes, transcription factors (TFs), polymerases, and recombinases, is suppressed. The question of how DNA-binding proteins gain access to their target sites \emph{in vivo} efficiently and robustly is one of the outstanding puzzles in chromatin biology.

Two different mechanisms for site exposure have been proposed: nucleosome translocation by thermal fluctuations or by ATP-dependent chromatin remodeling factors~\cite{Schiessel2001,Kulic2003,Vignali2000}, and transient unwrapping of nucleosomal DNA off the histone octamer surface~\cite{Polach1995,Anderson2002,Tims2011,Moyle-Heyrman2011}.
Proteins can utilize spontaneous DNA unwrapping to bind to their target sites, which would favor further destabilization of the histone-DNA complex and binding of additional proteins (Fig.~1A). Since partial unwrapping of nucleosomal DNA is energetically less costly than nucleosome translocation, it is likely to play a major role in numerous DNA-mediated processes. For example, nucleosome ``breathing'' governs transcription dynamics of RNA polymerase~\cite{Hodges2009}, and may provide an explanation for fast DNA repair by photolyases~\cite{Bucceri2006}.

Partial unwrapping of nucleosomal DNA and subsequent differential accessibility of nucleosome-covered protein-binding sites were observed in
single nucleosomes \cite{Adams1995,Polach1995,Anderson2000,Anderson2002,Miller2003,Li2004,
Li2005,Tims2011,Moyle-Heyrman2011}, di-nucleosomes~\cite{Engeholm2009}, and multi-nucleosome arrays~\cite{Poirier2008,Poirier2009}, and modeled computationally~\cite{Chou2003,Mirny2010,Teif2010,Teif2011,Teif2012,Prinsen2010}.
Recently, nucleosome dyad positions were mapped with single base-pair (bp) precision in \textit{S.cerevisiae}, resulting in a genome-wide collection of distances between neighboring nucleosomes~\cite{Brogaard2012}. Surprisingly, almost $40$\% of these inter-dyad distances are less than 147~bp, indicating that at least one nucleosome in the pair is partially unwrapped. Furthermore, there are distinct 10-11 bp periodic oscillations in the histogram of inter-dyad distances, suggesting a stepwise unwrapping mechanism consistent with the pattern of histone-DNA contacts in the crystal structure~\cite{Luger1997,davey:2002,richmond:2003}.

We present a rigorous statistical mechanics approach to nucleosome unwrapping which explains the observed genome-wide distribution of inter-dyad distances~\cite{Brogaard2012} as well as earlier experiments which probed differential accessibility of nucleosome-covered binding sites~\cite{Polach1995,Anderson2002} and studied nucleosome-induced cooperativity between DNA-binding factors~\cite{Adams1995,Moyle-Heyrman2011}.
Using this approach, we reproduce
both nucleosome occupancy and inter-dyad distance
profiles in the vicinity of transcription start sites (TSS).
The model neglects sequence specificity of histone-DNA
interactions but requires potential barriers at gene promoters, which
may be created by DNA-bound TFs and transcription pre-initiation
complexes (PICs).
Although sequence-specific nucleosome positioning is not crucial for explaining
many features of the distribution of inter-dyad distances, our framework allows us to predict both sequence-specific free energies of nucleosome formation and the unwrapping potential from paired-end high-throughput sequencing datasets.

\section{Results}

\subsection{Nucleosome unwrapping potential} \label{Inter-dyad distances}

We use a high-resolution \textit{in vivo} map of nucleosome dyad positions based on chemical modification of engineered histones and DNA backbone cleavage by hydroxyl radicals~\cite{Brogaard2012}.
With cleaved DNA segments sorted by molecular mass on the agarose gel and the $\sim 150-200 \text{ bp}$ fraction
sequenced using paired-end reads, the map provides a direct measurement of both
dyad positions and distances between adjacent dyads.
Although superior to methods based on micrococcal nuclease (MNase) digestion whose accuracy is affected by MNase sequences preferences and its tendency to over- or under-digest DNA~\cite{dingwall:1981,mcghee:1983,chung:2010}, the map is biased by unknown hydroxyl radical cutting preferences for two alternate sites at each DNA strand~\cite{Brogaard2012}.

A genome-wide histogram of inter-dyad distances shows that at least 38.7 \% of all nucleosomes are partially unwrapped (blue line in Fig.~1C).
In order to study the energetics of unwrapping, we introduce a simple model
based on the 10-11 bp periodic pattern of histone contacts with the minor groove of the nucleosomal DNA~\cite{davey:2002,richmond:2003} (Fig.~1B).
As DNA is peeled off each contact patch, its free energy increases because hydrogen bonds and favorable electrostatic contacts between histone side chains and the DNA phosphate backbone are lost.
However, once DNA breaks free from the contact patch, it may adopt multiple conformations, which allows it to increase its entropy and thus lower the total free energy.
The favorable entropic term grows with the extent of unwrapping until the next contact patch is reached, completing one cycle in the oscillatory energy profile. The oscillations are superimposed on a straight line whose slope reflects the average free energy cost per bp of histone-DNA contact loss.
Additional details of the potential construction can be found in SI Appendix, Model A.
The histone-DNA potential constructed in this way has no sequence specificity, in contrast to the free energy cost of bending DNA into a nucleosomal superhelix~\cite{morozov:2009}. To a good approximation, we expect sequence effects to average out in the genome-wide histogram of inter-dyad distances (this assumption is tested later).

Thus we aim to reproduce the observed distribution of inter-dyad distances with a model in which nucleosome energetics is sequence-independent but transient unwrapping is allowed.
We compute the conditional probability $P(c+d|c)$ of finding the nucleosome
dyad at bp $c+d$, given that the previous dyad is located at bp $c$ (Materials and Methods).
Since inter-dyad distances cannot be used to distinguish between symmetric and asymmetric unwrapping, we assume the former for simplicity.
The model is fit to the observed distribution of inter-dyad distances (SI Appendix).
The free parameters of the model include the amplitude of the oscillations, the slope of the free energy profile and $a_{\text{min(max)}}$, the minimum (maximum) effective length of the nucleosome particle (SI Appendix, Model A). The maximum extent of nucleosome unwrapping is controlled by
$a_{\text{min}}$, while $a_{\text{max}}$ is allowed to exceed $147 \text{ bp}$ in order to account for the effects of higher-order chromatin structure and linker histone deposition.
We also fit the relative frequency of hydroxyl radical DNA cleavage at the $-1$ position with respect to the nucleosome dyad (SI Appendix) and the chemical potential of histone octamers.
Our model reproduces both the overall shape and fine oscillatory structure of the observed inter-dyad distance distribution (Fig.~1C,D).
In contrast, models without unwrapping are unable to capture even the overall shape of the observed inter-dyad distribution (gray line in Fig.~1C).

\subsection{Higher-order chromatin structure and linker histone energetics} \label{higher-order}

The effective length of the particle found in the fit, $a_{\text{max}} = 163 \text{ bp}$, is greater than 147 bp, the length of the DNA in the nucleosome core~\cite{richmond:2003}.
Indeed, $a_{\text{max}} = 147 \text{ bp}$ is incompatible with the observed inter-dyad distribution (Fig.~S3A,B). The model is less sensitive to the value of $a_{\text{min}}$
because extensively unwrapped particles are energetically unfavorable and therefore are not frequently seen in the data.
The overall shape of the inter-dyad distribution is also sensitive to the slope of the energy profile in Fig.~1B, providing a robust estimate of the average nucleosome unwrapping energy (Fig.~S3C). The fitted value of the slope yields
$14.4 \text{ k}_{\text{B}} \text{T}$ for the average histone-DNA interaction energy in a fully wrapped nucleosome.

Thus the energy profile in Fig.~1B describes both DNA interactions with the histone octamer core (up to 73 bp from the dyad)
and the effects of higher-order chromatin structure, including, potentially, the attachment of Hho1p, the H1 linker histone of \textit{S.~cerevisiae}, to the DNA immediately outside of the nucleosome core~\cite{Zlatanova2008,Syed2010,Georgieva2012}.
Although Hho1p is less abundant in yeast than in higher eukaryotes, it is
involved in higher-order chromatin organization, including chromatin compaction in stationary phase~\cite{Schafer2008,Georgieva2012}.
Relatively little is known about the molecular mechanism of H1 binding: there is no consensus yet whether the binding is symmetric
or asymmetric, or even what the extent of the H1 footprint is~\cite{Zlatanova2008,Syed2010}. H1 binding and other factors that mediate chromatin folding
into higher-order structures cause linker lengths to be discretized~\cite{vanholde:1989,widom:1992}. Linker length discretization can be described by a periodic decaying two-body effective potential between neighboring nucleosomes, with the first minimum approximately 5 bp away from the nucleosome edge~\cite{vanholde:1989,wang:2008,Chereji2011a}.

\begin{figure}[t!]
\centerline{
\includegraphics[width=2.05in]{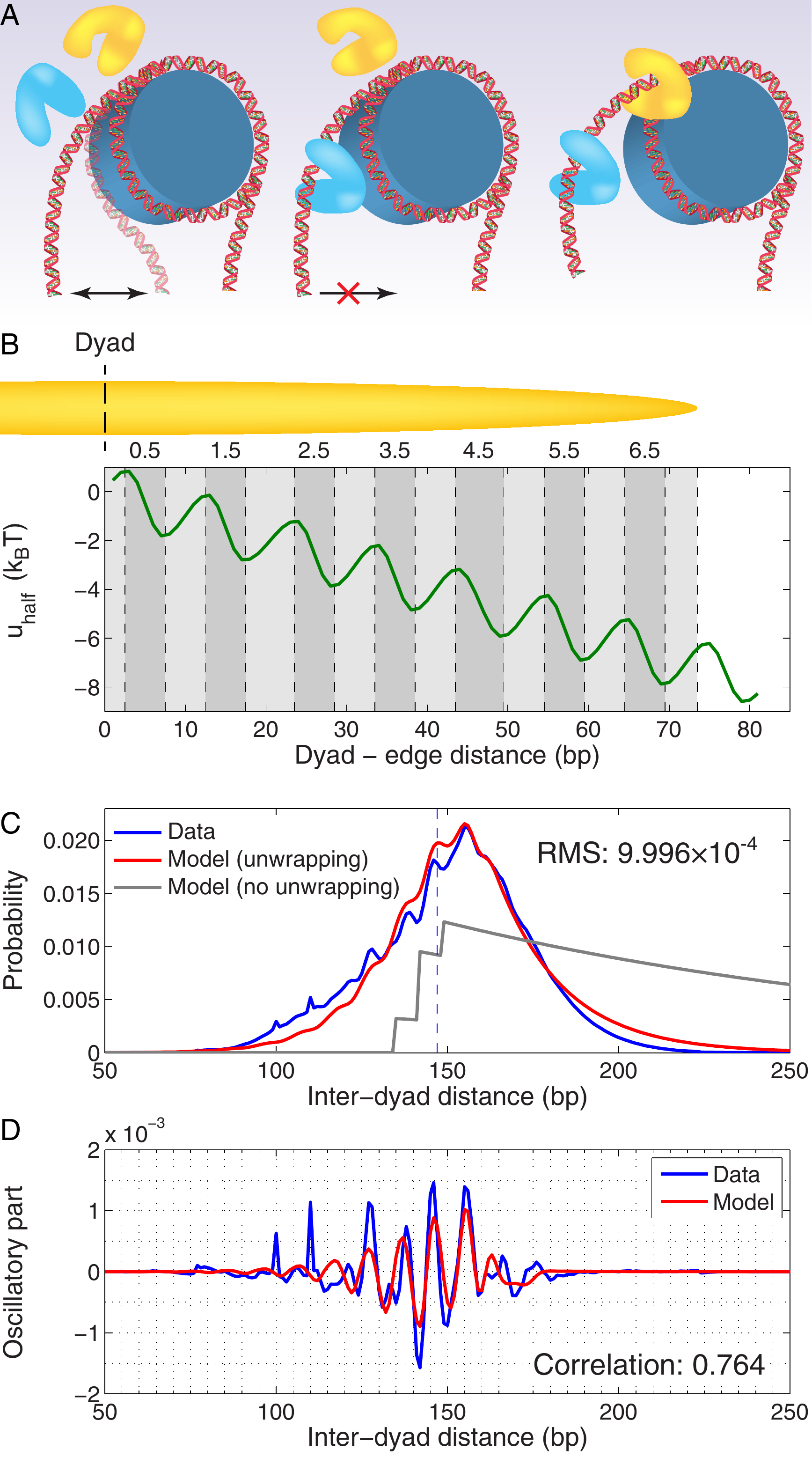}
}
\caption{\textbf{Genome-wide distribution of inter-dyad distances.}
(A) A nucleosome is a dynamic structure. Transient nucleosome unwrapping followed by factor binding prevents subsequent rewrapping, mediating further binding events.
(B) Nucleosome energy profile.
The energy of a nucleosome that covers $2x + 1$~bps is given by $u^{\mathrm{SI}}_\text{nuc} = 2 u_\textrm{half}(x)$ with symmetric unwrapping.
The minima and maxima of the energy landscape are based on the crystal structure of the nucleosome core particle~\cite{davey:2002,richmond:2003}. Dark gray bars show where the histone binding motifs interact with the DNA minor groove in the structure.
Light gray bars show where the DNA major groove faces the histones. The energy profile was obtained by a polynomial fit as described in SI Appendix, Model A.
(C) The inter-dyad distance distribution from a high-resolution nucleosome map~\cite{Brogaard2012} (blue),
and from the model with (red) and without unwrapping (gray). In the model without unwrapping, $a_\text{min} = a_\text{max} = 147$~bp and the fitting parameters are
$E_b$, $\mu$ and $f$ (SI Appendix, Model A).
RMS - total root-mean-square deviation between the model and the data.
(D) Oscillations in the observed (blue) and predicted (red) inter-dyad distance distributions, obtained
by subtracting a smooth background from the data and the model with unwrapping in (C). The smooth background was found by applying a Savitzky-Golay
filter of polynomial order 3 with 31 bp length (using the \texttt{sgolayfilt} function from the Signal Processing Toolbox of MATLAB).
Correlation refers to $r_{\text{osc}}$, the linear correlation coefficient between measured and predicted oscillations.
}
\label{Fig:InterDyad}
\end{figure}

Based on these observations, we have constructed two models for the energy profile outside of the nucleosome core region.
One model is a polynomial fit that extends the quasiperiodic profile of the unwrapping energy through another cycle (Fig.~1B; SI Appendix, Model A).
The depth and the position of the first minimum outside of the nucleosome core are additional free parameters.
As can be seen in Fig.~S4B, our fit robustly predicts the first minimum to be positioned 5-6 bp outside of the nucleosome core, in agreement with previous studies~\cite{vanholde:1989,wang:2008,Chereji2011a}.
The depth of this minimum is comparable to the depth of the unwrapping minima
(Fig.~S4, SI Appendix, Model A).

The other model represents the energy profile outside of the nucleosome core by a linear function (Fig.~S5A; SI Appendix, Model B).
The two free parameters are the slope and the length of the line, which are related to the H1-DNA interaction energy and the H1 footprint.
This model is motivated by a picture of the H1 histone being gradually detached from its DNA site immediately outside of the nucleosome core. This alternative scenario,
although likely oversimplified, can be used to check the sensitivity of our results toward a particular energy profile outside of the core region.
We find that the linear profile fits the overall shape of the inter-dyad distribution somewhat less well than the oscillatory one (compare RMS values in Figs.~1B and S5B), 
although the 10-11 bp periodic fine structure is reproduced in both cases (Figs.~1C, S5C). The optimal linear profile is 7 bp long, yielding a symmetric H1 footprint with two 7 bp half-sites (Fig.~S5D) and the H1-DNA interaction energy of $\approx 5 \text{k}_{\text{B}} \text{T}$ (Fig.~S5E). 

\subsection{Alternative models of nucleosome unwrapping} \label{alt-models}
Next we have tested the sensitivity of our fits to the functional form of the unwrapping free energy profile. Although our primary model follows nucleosome crystal
structures in creating a quasi-periodic energy profile with both 10 and 11 bp modes,
strictly periodic 10 or 11 bp sinusoidal profiles yield nearly the same quality of fit (Fig.~S6; SI Appendix, Models C,D).
Since the initial phase of the oscillations is not determined by the crystal structure anymore, it becomes another fitting parameter. The fitted initial phases in the 10 and 11-bp models make the periodic curves match the crystal structure further away from the dyad, where most of the observed unwrapping takes place (Fig.~S6A).
The phases diverge closer to the dyad, where they are not as strongly constrained by the data.
RMS deviation is less sensitive to the initial phase than $r_{\text{osc}}$, the linear correlation between predicted and observed oscillations
in the inter-dyad histograms (Fig.~S7). The primary peak in the dependence of $r_{\text{osc}}$ on the initial phase matches the crystal structure.
There is also a secondary peak corresponding to the $5 \text{ bp}$ shift in the unwrapping energy profile, which in turn leads to the $10 \text{ bp}$, in-phase
shift in the distribution of inter-dyad oscillations (Fig.~S7B).

Since the inter-dyad distance distribution has a distinct oscillatory component, it is not surprising that a purely linear model of unwrapping energy does not fit the data
as well, although it does match its overall shape (Fig.~S8A; SI Appendix, Model E). Less trivially, it was suggested on the basis of single-nucleosome
unzipping experiments that nucleosome unwrapping proceeds with 5-bp periodicity because histones interact with each DNA strand separately where
the DNA minor groove faces the histone octamer surface, creating two distinct contact ``subpatches''~\cite{Hall:2009}.
This single-molecule data was fit to a model with a step-wise unwrapping free energy profile~\cite{Forties:2011}.
Each step in the profile corresponds to breaking a point histone-DNA contact, and the steps occur every $5.25$ bp on average. 
We do not find any evidence for 5 bp periodicity of nucleosome unwrapping in the genomic data. Indeed, both 5 bp step-wise and 5 bp periodic
sinusoidal profiles fit the data poorly, about as well as the linear model (Fig.~S8B,C).
Even the 10-bp step-wise unwrapping profile, while clearly having the right periodicity, does not fit the data as well as the structure-based model
(Fig.~S8D). This observation suggests that the picture of gradual loss of favorable finite-range histone-DNA interactions
followed by gain in DNA conformational entropy is closer to reality than abrupt disruption of short-range histone-DNA contacts.
A direct comparison of single-molecule and genome-wide energy profiles is unfortunately obscured by the fact
that the reported single-nucleosome unzipping experiments are specific to the 601 nucleosome-forming sequence~\cite{thastrom:1999},
in contrast to our methodology which provides the average, sequence-independent picture of unwrapping energetics.

\begin{figure}[t!]
\centerline{
\includegraphics[width=3.46in]{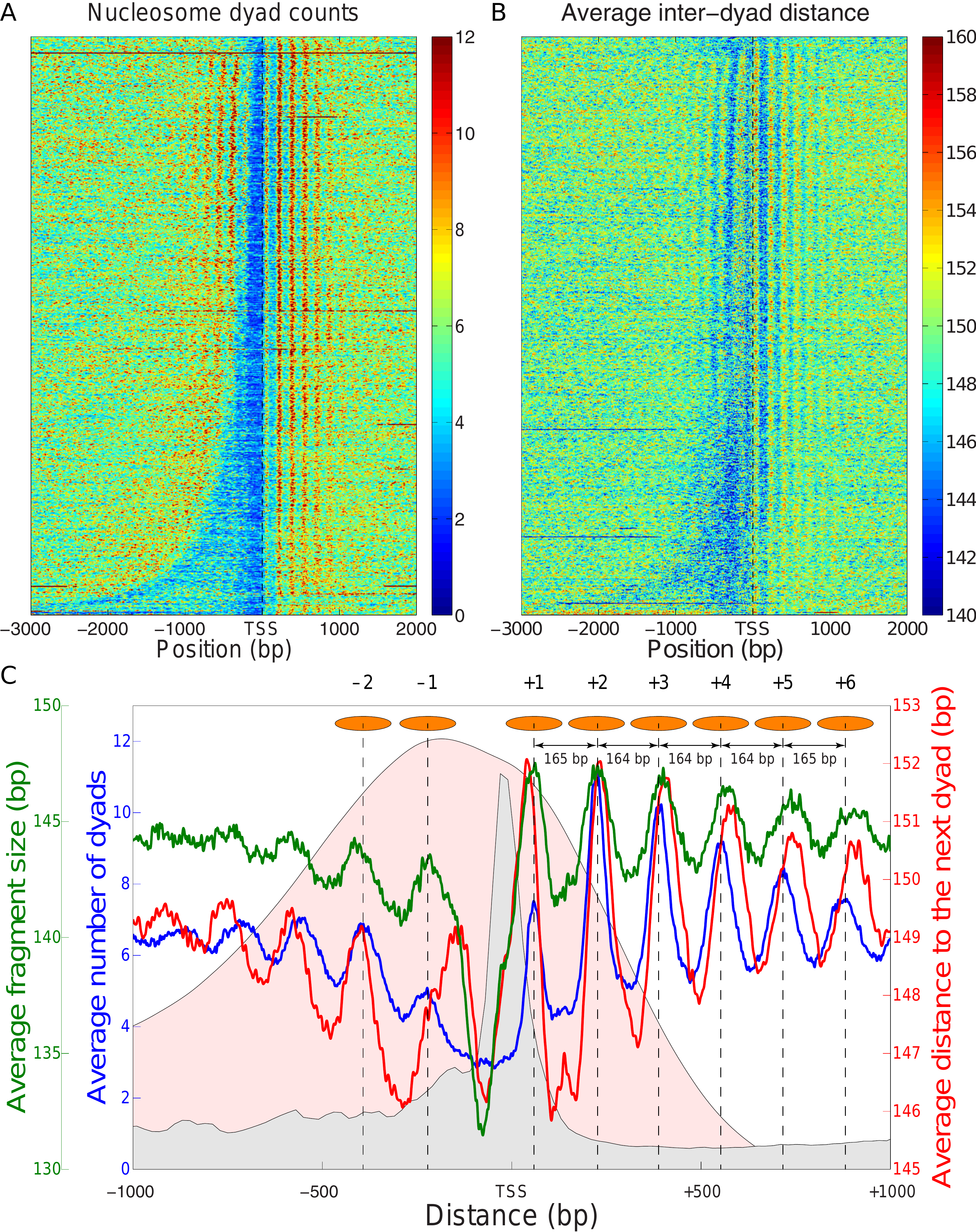}
}
\caption{\textbf{Nucleosome unwrapping in the vicinity of transcription start sites.}
(A) Distribution of nucleosome dyad counts~\cite{Brogaard2012} near the TSS.
4763 verified \textit{S.cerevisiae} open reading frames (ORFs)
were aligned by their TSS and sorted by promoter lengths. Each horizontal line corresponds to one ORF.
(B) Distribution of the average distance between neighboring dyads. For each bp, the distances between a dyad at that bp and all neighboring dyads were averaged.
ORFs are sorted as in (A). In (A) and (B), values at bps without dyads were obtained by interpolation, and
heatmaps were smoothed using a 2D Gaussian kernel with $\sigma = 3$ pixels.
(C) Data in (A), (B) and Fig.~S9A-C averaged over all genes. Blue: nucleosome dyad counts, red: average distance between neighboring dyads, green: average length of DNA-bound particles mapped by MNase digestion~\cite{henikoff:2011} (see Fig.~S9A for details). Curve with light gray background: combined occupancy of
9 PICs (TBP, TFIIA, TFIIB, TFIID, TFIIE, TFIIF, TFIIH, TFIIK, PolII)~\cite{rhee:2012},
curve with light pink background: average histone turnover rate~\cite{dion:2007}.
The peaks in the dyad count profile (blue) are marked with orange ovals representing nucleosomes, and peak-to-peak distances are shown.
}
\label{Fig:GenomicUnwrapping}
\end{figure}

\subsection{Genome-wide organization of nucleosome unwrapping states} \label{Unwrapping observations}

Fig.~2A, in which genes are sorted by the promoter length and aligned by the TSS, shows a canonical picture of nucleosomes depleted in promoters and well-positioned over coding regions~\cite{yuan:2005,mavrich:2008b}. Interestingly, promoter nucleosomes have shorter inter-dyad distances and are therefore more unwrapped (Fig.~2B). When averaged over all genes, the number of dyads at a given bp and the average inter-dyad distance at that bp
are strongly correlated (compare blue and red lines in Fig.~2C).
The profile of average inter-dyad distances is also correlated with the distribution of DNA fragment lengths in an MNase assay which mapped both nucleosomes and subnucleosome-size particles by paired-end sequencing (Fig.~S9A, green line in Fig.~2C)~\cite{henikoff:2011}.
The two profiles do not coincide completely because inter-dyad distances also depend on the distribution of linker lengths.
The observed behavior is opposite of the naive expectation that unwrapping increases with occupancy due to nucleosome crowding, but can be reproduced in a simple sequence-independent model in which nucleosomes phase off a potential barrier placed in the promoter region (Fig.~S10)~\cite{Chereji2011b}.

Partially unwrapped nucleosomes tend to have elevated histone turnover rates~\cite{dion:2007}, both around TSS and genome-wide (Figs.~2C,S9B,S9D). We find that nucleosomes at loci enriched in PICs~\cite{rhee:2012} are also more unwrapped (Figs.~2C,S9C). Finally,
inter-dyad distances tend to increase with the fraction of A/T nucleotides, indicating that nucleosomes occupying A/T-rich sequences have longer footprints genome-wide (Fig.~S9E).
We note that it is misleading to equate inter-nucleosome distances with peak-to-peak distances in the average profile of nucleosome dyad counts (blue line in Fig.~2C).
The peak-to-peak distances are 164-165 bp, while the average inter-dyad distance for
the nucleosomes in the [TSS,TSS+1000] region is 149.6 bp.
Thus nucleosome unwrapping is much more common than could be predicted by mapping single-nucleosome positions alone.

\begin{figure}[t!]
\centerline{
\includegraphics[width=2.5in]{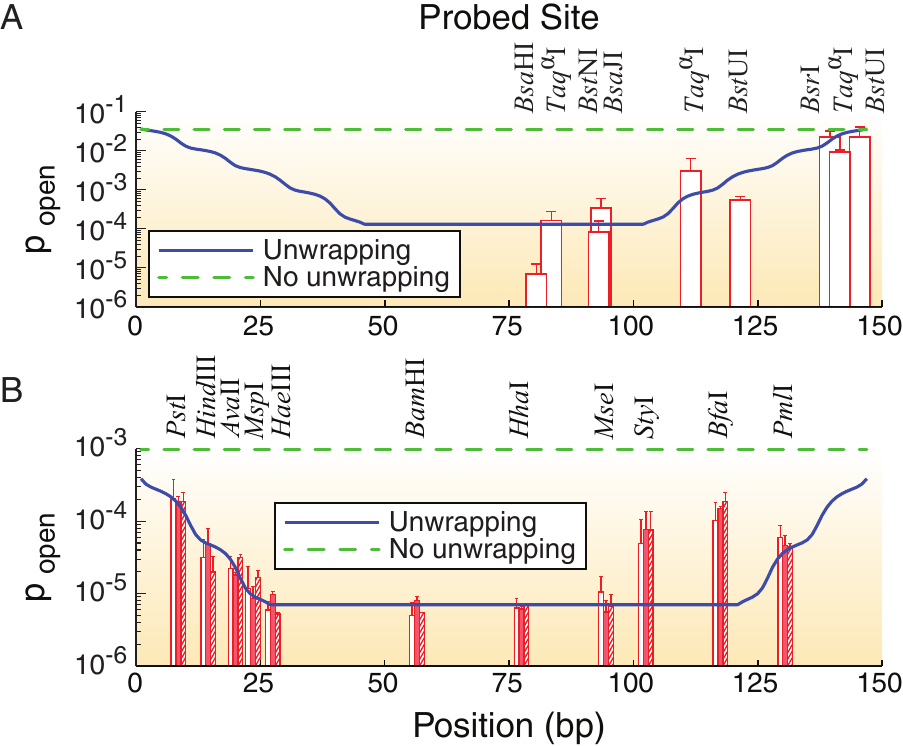}
}
\caption{\textbf{Probability of binding site exposure within a nucleosome.}
The solid blue and dashed green lines represent model predictions with and without unwrapping, respectively. In the latter case, $a_{\text{min}} = a_{\text{max}} = 147$~bp and all other parameters are adopted from the model with unwrapping.
The dyad is fixed at bp 74.
(A) Restriction enzyme sites inserted into the 5S rRNA sequence at locations indicated by the centers of vertical red bars~\cite{Polach1995}.
(B) Restriction enzyme sites inserted into the 601 sequence at locations indicated by the centers of the vertical red bars in the middle of each group~\cite{Anderson2002}.
Each group of three bars corresponds to independent measurements in which the 601 sequence was flanked by different DNA sequences.   
In (A) and (B), the height of each bar is the equilibrium constant for site exposure averaged over multiple experiments (error bars show standard deviation).
}
\label{Fig:Accessibility}
\end{figure}

\subsection{Accessibility of nucleosomal DNA to factor binding} \label{Accessibility problem}

Partial unwrapping of nucleosomal DNA results in differential accessibility of factor binding sites with respect to their position inside the nucleosome:
sites on the edges are more accessible than those closer to the dyad. In contrast, all-or-none nucleosome formation should not be sensitive to the binding site position -- a nucleosome, once unfolded, liberates its entire site. 
Polach and Widom~\cite{Polach1995} studied differential accessibility of 6 restriction enzymes
to their target sites.
The sites were placed at various positions throughout the 5S rRNA nucleosomal sequence (Fig.~3A).
A later study used the 601 sequence and an extended set of 11 restriction enzymes (Fig.~3B)~\cite{Anderson2002}.
These studies measured equilibrium constants for site exposure $K_\text{eq}^\text{conf}$, which are related to the probability for a site to be accessible
for binding: $p_\text{open} = K_\text{eq}^\text{conf} / (1 + K_\text{eq}^\text{conf}) \approx K_\text{eq}^\text{conf}$~\cite{Prinsen2010}. 

We use our crystal structure-based unwrapping model (Fig.~1B; SI Appendix, Model A) to fit the data on site accessibility~\cite{Polach1995,Anderson2002}.
Here the system consists of a single nucleosome and asymmetric unwrapping is allowed.
We assume that a site becomes accessible for the enzyme only after an additional number of bps, $d$, have been unwrapped
from the histone octamer surface~\cite{Prinsen2010}. We also assume that once the dyad is unwrapped from either end, the entire nucleosome is unfolded.
Then the probability for a binding site to be accessible is given by
\[
p_\textrm{open}(x) = 
\left\{
\begin{array}{l l}
1 - \Occ_\textrm{nuc}(x + d) &\textrm{for }x < x_\textrm{d} - d\\
1 - \Occ_\textrm{nuc}(x_\textrm{d}) & \textrm{for }x_\textrm{d} - d \leq x \leq x_\textrm{d} + d\\
1 - \Occ_\textrm{nuc}(x - d) & \textrm{for }x > x_\textrm{d} + d,
\end{array}\right.
\]
where $x \in [1,147] \text{ bp}$, $x_\textrm{d} = 74 \text{ bp}$ is the position of the dyad, and the nucleosome occupancy is given by Eq.~\eqref{Eq:Occ}.

Besides $d$, the fitting parameters of the model are the overall slope of the binding energy $\epsilon$ and the histone chemical potential $\mu$ (all other parameters are as in SI Appendix, Model A, with the exception of $a_{\text{min}} = 1 \text{ bp}$, $a_{\text{max}} = 147 \text{ bp}$). For the 5S rRNA measurements~\cite{Polach1995}, we obtain $\epsilon^\text{5S} = -0.13\text{ k}_\text{B}\text{T/bp}$, $\mu = -17.5\text{ k}_\text{B}\text{T}$, and $d = 23\text{ bp}$. For the 601 measurements~\cite{Anderson2002}, $\epsilon^\text{601} = -0.16\text{ k}_\text{B}\text{T/bp}$, $\mu = -16.4\text{ k}_\text{B}\text{T}$, and $d = 45\text{ bp}$. As expected, the nucleosome formation energy of the 601 sequence is $147 \times (\epsilon^\text{5S} - \epsilon^\text{601}) = 4.4 \text{ k}_\text{B}\text{T}$ more favorable than that of the 5S sequence, in agreement with the experimentally measured difference of 4.9 $\text{ k}_\text{B}\text{T}$~\cite{Thastrom2004}. The nucleosome formation energy of the 601 sequence is $24.1 \text{ k}_\text{B}\text{T}$, close to the $23.8 \text{ k}_\text{B}\text{T}$ estimate made on the basis of 601 unzipping experiments~\cite{Forties:2011}. Interestingly, the 601 DNA has to unwrap more extensively past the binding site to allow access to restriction enzymes.

Overall, our model reproduces the observed differential accessibility of restriction enzyme binding sites with respect to the nucleosome dyad (Fig.~3).
The only outliers are \emph{Sty}I and \emph{Bfa}I binding sites in the 601 series which were not used in the fit and
which, if unwrapping proceeds from the ends, cannot be more open than the \emph{Pml}I site located further away from the dyad.

\subsection{Nucleosome-induced cooperativity} \label{Nuc_coop}
If multiple biding sites reside within a single nucleosome, binding of one factor makes the other sites more accessible, in a phenomenon known as
nucleosome-induced cooperativity~\cite{Adams1995,Miller2003,Mirny2010}. The cooperativity disappears in the absence of nucleosomes and reduces in extent with the distance between consecutive sites~\cite{Adams1995}. Moreover, the cooperativity is not observed
if the two sites are on the opposite sides of the nucleosome dyad~\cite{Moyle-Heyrman2011}.

We can use our model of nucleosome unwrapping (SI Appendix, Model A with $a_{\text{min}} = 1 \text{ bp}$, $a_{\text{max}} = 147 \text{ bp}$) to capture all these aspects of nucleosome-induced cooperativity (Fig.~4).
Specifically, for sites located more than $40 \text{ bp}$ away
from the dyad site accessibility is strongly enhanced if DNA unwrapping is allowed (Fig.~4A). Interestingly, cooperativity between two TFs bound on the same
side of the dyad is observed both with and without unwrapping (Fig.~4B).
However, without unwrapping it is impossible to show that binding on the opposite sides of the dyad is not cooperative, as observed in experiments~\cite{Moyle-Heyrman2011} (Fig.~4C).
Furthermore, the decrease of cooperativity with distance~\cite{Adams1995} cannot be reproduced (Fig.~4D).
Thus modeling transient nucleosome unwrapping is necessary for understanding how TFs and other DNA-binding proteins gain access to their nucleosome-covered sites.

\subsection{Sequence-dependent nucleosome positioning and unwrapping}
Here we test the assumption that sequence specificity of nucleosome formation may be neglected when inferring nucleosome unwrapping potential from genomic data.
In general, the total free energy $u_\text{nuc}(k,l)$ of a nucleosome occupying bps
$k, \ldots ,l$ is a sum of a sequence-independent term $u^{\mathrm{SI}}_\text{nuc}(k,l)$ describing contacts between histone side chains and the DNA phosphate backbone (Fig.~1B), and a sequence-dependent term $u^{\mathrm{SD}}_\text{nuc}(k,l)$ describing DNA bending into the nucleosomal superhelix~\cite{morozov:2009}. Our approach (Eq.~\eqref{Eq:u}, SI Appendix) can be used to infer both contributions from high-throughput maps of nucleosome positions.
Since the number of different unwrapped species may be as high as several thousand
depending on the maximum extent of unwrapping and the assumption of unwrapping symmetry, we expect robust inference to require levels of read coverage that are not currently available, especially in higher eukaryotes. Furthermore, resolution of MNase-based nucleosome maps is insufficient for capturing the 10-11 bp periodicity of the unwrapping potential.
In the absence of high-resolution, high-coverage experimental data, we have tested our ability to predict nucleosome unwrapping energetics using a realistic model system.

Specifically, we assumed that $u^{\mathrm{SD}}_\text{nuc}$ depends only on the number of mono- and dinucleotides in the nucleosomal DNA~\cite{Locke2010} (SI Appendix).
The length of nucleosomal DNA ranges from $a_{\text{min}} = 74 \text{ bp}$ to $a_{\text{max}} = 147 \text{ bp}$; all other parameters of $u^{\mathrm{SI}}_\text{nuc}$ are as in SI Appendix, Model A.
Using Eq.~\eqref{Eq:n1_2} with $\mu = -13\text{ k}_\text{B}\text{T}$,
we computed the exact nucleosome distribution $n_1^\text{nuc} (k,l)$ for the \textit{S.cerevisiae} chromosome I.
We sampled paired-end nucleosomal reads $[k,l]$ from $n_1^\text{nuc} (k,l)$ until a desired level of read coverage (mean number of reads starting at a bp) was reached.
From this finite sample, we constructed a histogram of nucleosome lengths $P(\Delta)$ and used it to optimize the parameters of $u^{\mathrm{SI}}_\text{nuc}$ (in each step of the fitting procedure, $u^{\mathrm{SI}}_\text{nuc}$ alone was used to predict $n_1^\text{nuc}$, which in turn gave $P(\Delta)$). Next, we used Eq.~\eqref{Eq:u} to predict $u_{\text{nuc}} (i,j)$, subtracted $u^{\mathrm{SI}}_\text{nuc}$ from it (assuming that the dyad is at the mid-point of each particle), and fit $u^{\mathrm{SD}}_\text{nuc}$ to the rest. Finally, $u^{\mathrm{SI}}_\text{nuc} + u^{\mathrm{SD}}_\text{nuc}$ were used to compute $\tilde{n}_1^\text{nuc}$, which was then compared with the exact profile $n_1^\text{nuc}$.

\begin{figure}[th!]
\centerline{
\includegraphics[width=3.5in]{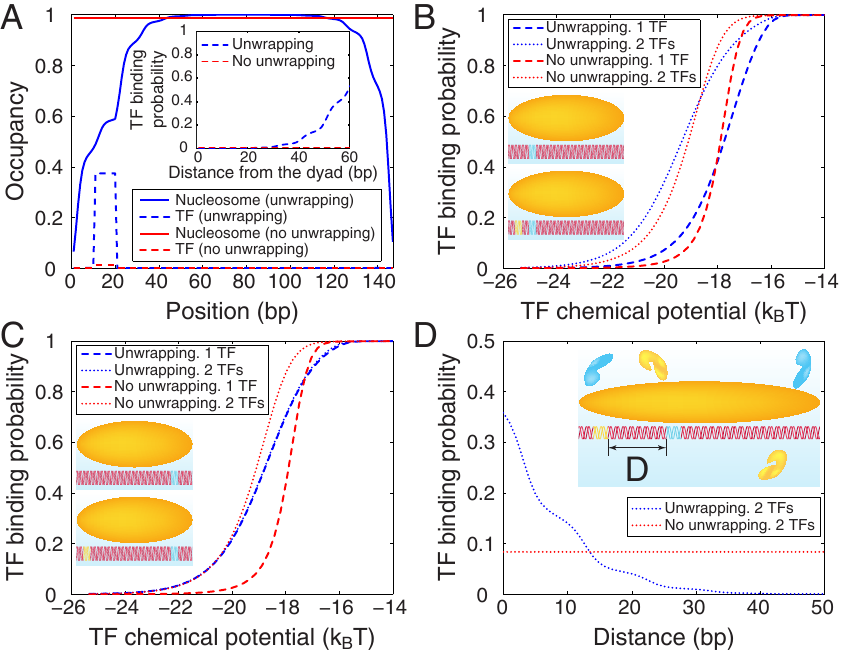}
}
\caption{\textbf{Modes of nucleosome-induced cooperativity.}
(A) TF and nucleosome occupancy with and without unwrapping. The TF binding site occupies bps 11--20.
Inset: TF binding probability as a function of the distance between the nucleosome dyad and the proximal edge of the TF site, with and without unwrapping.
(B) TF titration curves for one TF site vs. two TF sites located on the same side of the dyad.
Site 1 occupies bps 11--20, site 2 occupies bps 31--40.
Inset: Binding site locations.
(C) Same as (B), but with the two TF sites located on the opposite sides of the dyad.
Site 1 occupies bps 11--20, site 2 occupies bps 117--126.
Inset: Binding site locations.
(D) Nucleosome-induced cooperativity as a function of the distance between two TF binding sites.
The binding probability of the second TF is shown. Site 1 occupies bps 11-20, while the position of the second site is variable.
Inset: Definition of the distance between the two binding sites.
In all panels, free energy of a fully wrapped nucleosome is $-\log(10^9)\text{ k}_\text{B}\text{T}$;
histone chemical potential is $\log(10^{-6})\text{ k}_\text{B}\text{T}$; TF binding energy is $-\log(10^{10})\text{ k}_\text{B}\text{T}$ to cognate sites,
and $-\log(10^{6})\text{ k}_\text{B}\text{T}$ to all other sites; TF chemical potential is $\log(10^{-9})\text{ k}_\text{B}\text{T}$ unless varied.
Asymmetric unwrapping is allowed; in the model without unwrapping, $a_{\text{min}} = a_{\text{max}} = 147$~bp and all other parameters are the same as in the model with unwrapping.
}
\label{Fig:CooperativeBinding}
\end{figure}

We find that we are able to reproduce the unwrapping potential even if nucleosome formation is sequence-specific (Fig.~S11A). However, the overall slope of the potential is overpredicted because the histogram of particle lengths is affected by well-positioned nucleosomes which have negative sequence-dependent energies. The average of these energies biases the slope. Nucleosome occupancy predicted using $u^{\mathrm{SI}}_\text{nuc} + u^{\mathrm{SD}}_\text{nuc}$ yields a reasonable correlation with the exact profile even at 1 read per bp, although at least 10 reads per bp are required to reproduce dyad positions (Fig.~S11B).

\section{Discussion and Conclusions}
Since DNA wrapped in nucleosomes is sterically occluded, nucleosome formation prevents access of DNA-binding factors such as TFs or polymerase subunits to their cognate sites.
Transient DNA unwrapping off the histone octamer surface, extensively studied in model systems~\cite{Adams1995,Polach1995,Anderson2000,Anderson2002,Miller2003,Li2004,
Li2005,Tims2011,Moyle-Heyrman2011,Engeholm2009,Poirier2008,Poirier2009}, can facilitate such binding events but its importance on the genomic scale has been unclear. MNase-based maps of
nucleosome positions employing paired-end sequencing have identified numerous subnucleosome-size particles~\cite{henikoff:2011,Cole2011}. However, it is impossible to deduce from these
experiments which particles correspond to unwrapped nucleosomes. In addition, there is a concern that DNA fragments may have been under- or overdigested by MNase.

These challenges have been overcome in a recent experiment in which chemically modified histones are used to map nucleosome dyads with single-bp resolution~\cite{Brogaard2012}. Used in conjunction with paired-end sequencing, the experiment provides information about both dyad positions and the distances between neighboring dyads. The latter is especially important since it probes the two-particle distribution in the same cell. The histogram of inter-dyad
distances shows that at least 38.7\% of all \textit{S.cerevisiae} nucleosomes are partially 
unwrapped. Gene promoters are enriched in such nucleosomes, which also have higher histone turnover rates. Over coding regions, the distribution of wrapped DNA lengths oscillates in 
phase with the nucleosome occupancy profile, so that more stable nucleosomes are also more wrapped.

We have developed a statistical mechanics description of nucleosome arrays which allows DNA unwrapping while rigorously treating steric exclusion and sequence specificity of nucleosome formation. We have shown that prominent 10-11 bp periodicity in the distribution of inter-dyad distances~\cite{Brogaard2012} can be explained
using an unwrapping energy profile based on the pattern of histone-DNA contacts in the
nucleosome crystal structures. We were able to rule out alternative scenarios of step-wise unwrapping and 5 bp periodicity of the unwrapping process. Furthermore, fitting the  distribution of inter-dyad lengths required accounting for linker length discretization, commonly thought to be imposed by chromatin fiber formation~\cite{widom:1992}.

Our model yields estimates of nucleosome unwrapping energies consistent with previous biophysical experiments, and accounts for single-nucleosome observations which show that nucleosome-covered binding sites closer to the edge of the nucleosome are more easily accessible to DNA-binding factors and that binding of the first factor enhances binding of subsequent factors on the same side of the dyad. Finally, our approach reproduces the periodic distribution of wrapped DNA lengths in the vicinity of TSS if potential barriers are placed in the promoters. The barriers may be created \textit{in vivo} by PICs~\cite{rhee:2012}, other DNA-bound factors, and chromatin remodelers.
The extent of nucleosome unwrapping in the yeast genome suggests that its treatment should be included in all future models of nucleosome positioning and chromatin energetics.

\section{Methods}
\subsection{Statistical mechanics of DNA-bound particles}
Here we outline the exact theory for $T$ molecular species simultaneously interacting with one-dimensional DNA
(details and extensions are provided in SI Appendix). The DNA-bound particles are subject to steric exclusion and may also be partially unwrapped.
Let $u_s(k,l)$ $(s \in \{1,2,\ldots,T\})$ denote the binding energy of a particle of type $s$ attached to bps $k, \ldots ,l$.
One can show that the one-body distribution of particles of type $t$ is given by
\begin{equation} \label{Eq:n1_2}
	n_1^t(k,l) = \frac{1}{Z} Z^-(k) \la t,k \ve z \ve t,l \ra Z^+(l),
\end{equation}
where $Z$ is the grand canonical partition function, $\la t,k \ve z \ve s,l \ra = e^{\beta [\mu_s - u_s(k,l)]} \delta_{t,s}$, $\beta = 1/k_BT$ is the inverse temperature,
$\mu_s$ is the chemical potential of particles of type $s$, and $\delta_{t,s}$ is the Kronecker symbol. If the DNA length is $L$,
vectors $\ve t,k \ra$ are $TL$-dimensional with a 1 at position $(t-1)L + k$ and 0 everywhere else.
Similarly, the nearest-neighbor two-body distribution function is given by
\begin{equation} \label{Eq:n2_main}
        \overline{n}_2^{t,s}(i,j; k,l) = \frac{1}{Z} Z^-(i) \la t,i \ve z \ve t,j \ra 
        \Theta(k-j) \la s,k \ve z \ve s,l \ra Z^+(l),
\end{equation}
where $\Theta$ is the Heaviside step function, and $Z^-(k)$ and $Z^+(k)$ are partial partition functions for the DNA segments [1,k) and (k,L], respectively. $Z^-(k)$, $Z^+(k)$, and $Z$ can be computed recursively, as shown in SI Appendix.
Using Eq.~\eqref{Eq:n1_2}, one can find particle occupancy:
\begin{equation}
\Occ_t(i) = \sum_{k=i-a^t_\text{max}+1}^i \sum_{l=\max(i, k+a^t_\text{min}-1)}^{k+a^t_\text{max}-1} n_1^t(k,l), \label{Eq:Occ}
\end{equation}
where $a^t_\text{min}$ ($a^t_\text{max}$) is the minimum (maximum) length
of a particle of type $t$.

The inverse problem of predicting DNA binding energies from one-particle distributions can also be solved:
\begin{equation}
\beta \left[ u_t(i,j) - \mu_t \right] = - \ln \left[ \frac{n_1^t(i,j) Z}{Z^{-}(i) Z^{+}(j)} \right], \label{Eq:u}
\end{equation}
where $Z, Z^{-}, Z^{+}$ are again found recursively using only one-particle distributions $n_1^t(i,j)$ as input (SI Appendix).
Eq.~\eqref{Eq:u} enables us to disentangle intrinsic sequence preferences and unwrapping energetics from steric effects.

\subsection{Model of inter-dyad distances}

To predict the distribution of inter-dyad lengths, we compute the conditional probability of having a nucleosome with the dyad at bp $c+d$,
given that the adjacent upstream nucleosome has the dyad at bp $c$:
\begin{equation}
P(c+d|c) = \frac{N_2(c,c+d)}{N_1(c)}, \label{Eq:CondProb}
\end{equation}
where the probability distributions of the nucleosome centers can be computed using Eqs.~\eqref{Eq:n1_2} and \eqref{Eq:n2_main} for a single particle type:
\begin{align*}
N_1(c) &= \sum_{\Delta_1}n_1^\text{nuc}(c-\Delta_1, c+\Delta_1),\\
N_2(c,c+d) &= \sum_{\Delta_1,\Delta_2}\overline{n}_2^{\text{nuc},\text{nuc}}(c-\Delta_1, c+\Delta_1; c+d-\Delta_2, c+d+\Delta_2).
\end{align*}
Here, $2 \Delta_{1,2} + 1$ are lengths of the particles centered at bp $c$ and $c+d$, respectively.
To estimate $P(c+d|c)$, we use $c = 5$~kbp and a box of length $L = 10$~kbp, so that the boundaries of the box are far away.
We ignore nucleosome sequence specificity, convolve $P(c+d|c)$ with a kernel to account for site-specific chemical cleavage bias (SI Appendix), and fit model parameters
to reproduce the observed distribution of inter-dyad distances.

\section*{Acknowledgments}
We thank Leonid Mirny for insightful discussions.
This research was supported by National Institutes of Health (R01 HG004708 to
A.V.M.). A.V.M. is an Alfred P. Sloan Research Fellow.
\newpage
\part*{\LARGE Supplementary Information}
\setcounter{section}{0}
\section{Supplementary Methods}

\renewcommand{\thefigure}{S\arabic{figure}}
\renewcommand{\theequation}{S\arabic{equation}}
\setcounter{figure}{0}
\setcounter{equation}{0}

\subsection{Direct problem: Matrix solution} \label{section:DirectProb}

\subsubsection{Single particle type} \label{subsection:DP_MatForm_1Sp}
We consider a problem of mutually interacting particles (one-dimensional rods) that can be reversibly adsorbed to a one-dimensional lattice of $L$ sites [here, DNA base pairs (bps)].
In order to model partial unwrapping of nucleosomal DNA off the surface of histone octamers, we allow the particles to cover a variable number of bps between $a_\text{min}$ and $a_\text{max}$. We assume that the particles cannot overlap while they are attached to the lattice.
This is implemented using hard-core interactions between adjacent particles.
There are also hard walls at the ends of the lattice so that particles are prevented from running off it.
In addition, we allow generic two-body interactions between nearest-neighbor particles.

The attachment of a particle to the DNA modifies the total energy of the system in a sequence-specific manner.
Physically, the binding energy may have contributions from DNA bending, electrostatic interactions, hydrogen bond formation, van der Waals contacts, etc.
Thus a particle which covers bps $k, k+1, \ldots ,l$ has a total one-body binding energy $u(k,l)$.
Note that for pairs of coordinates $(k,l)$ such that $l-k+1>a_\text{max}$ or $l-k+1<a_\text{min}$, $u(k,l)=\infty$, because all particles must have the
length between $a_\text{min}$ and $a_\text{max}$ bps. The theory presented below is valid for arbitrary binding energies $u(k,l)$.

\begin{figure}[t]
  \begin{center}
  \includegraphics[width=0.6\textwidth]{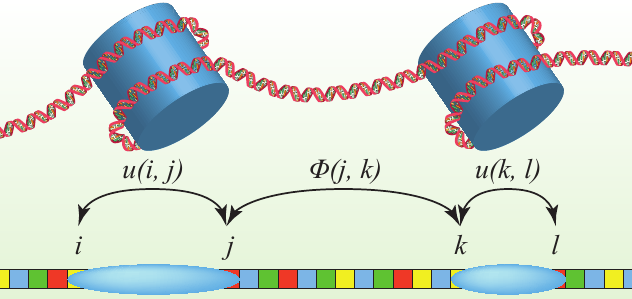}
  \end{center}
  \caption{Schematic illustration of one-body and two-body potentials in a multi-nucleosome 
system. Nucleosomes may be partially unwrapped, resulting in variable DNA footprints.
}
\label{Fig:Nucs}
\end{figure}

Let $\Phi(j,k)$ be the two-body interaction between a pair of nearest-neighbor particles which cover base pairs $\ldots, j-1, j$ and $k, k+1, \ldots$ ($k>j$) (Fig. \ref{Fig:Nucs}).
In the case of nucleosomes, such interactions may be used to account for the effects of higher-order chromatin structure~\cite{Chereji2011b}.
Although we do not focus on two-body interactions in this work, they are included below for the sake of generality.
We impose
\[
\Phi(j,k)= \left\{ \begin{array}{cl}
\infty & \text{if }k \leq j,\\
V(k-j-1) & \text{if }k>j, \end{array} \right.
\]
where $V(d)$ is an arbitrary interaction potential which depends only on the linear distance $d$ between two neighboring particles.

For a fixed number of particles attached to the DNA, $N$, the canonical partition function is
\begin{multline} \label{eq:QN}
	Q_N=\sum_{\{i_n = 1 \ldots L \}_{n\in \{1,\ldots,2N\}}} e^{-\beta u(i_1, i_2)} e^{-\beta \Phi(i_2, i_3)} e^{-\beta u(i_3, i_4)} \ldots \\
					                  \times e^{-\beta u(i_{2N-3}, i_{2N-2})} e^{-\beta \Phi(i_{2N-2},i_{2N-1})} e^{-\beta u(i_{2N-1}, i_{2N})},
\end{multline}
where $\beta = 1/k_B T$ is the inverse temperature ($k_B$ is the Boltzmann constant). Note that with our definitions of one-body energies, two-body interactions and hard-wall boundary conditions, only legitimate configurations of non-overlapping particles will contribute to Eq.~\eqref{eq:QN}.

In order to simplify the notation, we introduce two $L \times L$ matrices:
\begin{align*}
	\la k \ve e \ve l\ra &= e^{-\beta u(k,l)}, \\
	\la k \ve w \ve l\ra &= e^{-\beta \Phi(k,l)}.
\end{align*}

Here $\la k \ve M \ve l \ra$ represents the element of matrix $M$ in row $k$ and column $l$.
$\ve l \ra$ is a column vector of dimension $L$ with 1 at position $l$ and 0 everywhere else, and
$\la k \ve$ is a row vector with 1 at position $k$ and 0 otherwise.

Let $\ve J \ra$ be a vector of dimension $L$ with 1 at every position. Eq.~\eqref{eq:QN} gives
\[
	Q_N=\begin{cases} 
							\la J\ve (e w)^{N-1} e \ve J\ra &\text{if $N \geq 1$,} \\
							1 &\text{if $N=0$.} 
					\end{cases}
\]

If the particles are allowed to attach and detach from the lattice, the system has a variable number of particles, and the grand-canonical partition function is
\begin{align}
	Z &= \sum_{N=0}^{N_\text{max}} e^{\beta N \mu} Q_N \notag \\
		&= 1 + \sum_{N=1}^{N_\text{max}} \la J \ve (z w)^{N-1} z \ve J \ra \notag \\
		&= 1 + \sum_{M=0}^{\infty} \la J \ve (z w)^{M} z \ve J \ra \notag \\
		&= 1 + \la J \ve (I-z w)^{-1} z \ve J \ra, \label{Eq:Z}
\end{align}
where $\mu$ is the chemical potential, $N_\text{max}$ is the maximum number of particles that can fit on $L$ bp,
$I$ is the identity matrix, and $\la k \ve z \ve l\ra = e^{\beta [\mu-u(k,l)]} \equiv \zeta(k,l)$.
Note that all particle configurations with $N > N_\text{max}$ do not contribute to $Z$,
allowing us to extend the upper limit from $N_\text{max}$ to $\infty$.

From the partition function, we can compute $s$-particle distribution functions
(see the chapter by Stell in \cite{Frisch1964}):
\begin{align*} 
	n_1(k,l) &= \frac{\zeta(k, l)}{Z} \frac{\delta Z}{\delta \zeta(k, l)},\\
	n_2(i,j;k,l) &= \frac{ \zeta(i, j) \zeta(k, l)}{Z} \frac{\delta^2 Z}{
	\delta \zeta(i, j) \delta \zeta(k, l)},
\end{align*}
and in general
\[
n_s(i_{1L},i_{1R}; \ldots; i_{sL}, i_{sR}) = \frac{\zeta(i_{1L}, i_{1R}) \ldots \zeta(i_{sL}, i_{sR})}{Z} \frac{\delta^s Z}{\delta \zeta(i_{1L}, i_{1R}) \ldots \delta \zeta(i_{sL}, i_{sR})}.
\]

Using these relations, we obtain the one-particle distribution
\begin{equation} \label{Eq:n}
	n_1(k,l) = \frac{1}{Z} \la J \ve (I - z w)^{-1} \ve k \ra \la k \ve z \ve l \ra \la l \ve (I - w z)^{-1} \ve J \ra,
\end{equation}
and the two-particle distribution
\begin{equation} \label{Eq:n2}
	n_2(i,j;k,l) = \frac{1}{Z} \la J \ve (I - z w)^{-1} \ve i \ra \la i \ve z \ve j \ra \la j \ve w (I - z w)^{-1} \ve k \ra \la k \ve z \ve l \ra \la l \ve (I- w z)^{-1} \ve J \ra.
\end{equation}
In particular, the nearest-neighbor two-particle distribution is given by
\[
	\overline{n}_2(i,j;k,l) = \frac{1}{Z} \la J \ve (I - z w)^{-1} \ve i \ra \la i \ve z \ve j \ra \la j \ve w \ve k \ra \la k \ve z \ve l \ra \la l \ve (I- w z)^{-1} \ve J \ra.
\]

Eqs.~\eqref{Eq:n} and \eqref{Eq:n2} allow an obvious interpretation.
To find the probability of starting a particle covering positions from $k$ to $l$ [Eq.~\eqref{Eq:n}], we have to add statistical weights
of all the configurations that contain a particle at that position, and divide the resulting sum by the partition function.
Similarly, in order to find the probability of a pair of particles, one covering positions $i$ to $j$ and the other covering positions $k$ to $l$ [Eq.~\eqref{Eq:n2}],
we need to sum statistical weights of all the configurations containing that pair of particles, and divide by the partition function.

With the one-particle distribution, we can define the occupancy of a bp $i$ as the probability of finding that bp in contact with any particle.
In other words, we need to sum the probabilities of all configurations in which particles cover bp $i$:
\[
\operatorname{Occ}(i) = \sum_{k=i-a_\text{max}+1}^i \sum_{l=\max(i, k+a_\text{min})}^{k+a_\text{max}-1} n_1(k,l).
\]
Note that $1 - \operatorname{Occ}(i)$ is the probability that bp $i$ is not covered by any particles.

\subsubsection{Multiple particle types} \label{subsection:DP_MatForm_multSp}
\begin{figure}[t]
  \begin{center}
  \includegraphics[width=0.417\textwidth]{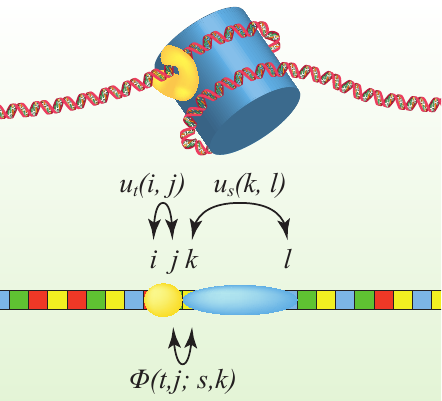}
  \end{center}
  \caption{Schematic illustration of one-body and two-body potentials in a system with multiple particle types. The model allows all particles to be in multiple stages of unwrapping. In practice, we allow nucleosomes to be partially unwrapped but the transcription factors (TFs) always have fixed DNA footprints. 
}
\label{Fig:NucsTF}
\end{figure}

The above formalism can be easily extended to the case in which $T$ types of particles can attach to the one-dimensional lattice.
Let the binding energy of a particle of type $t \in \{1, \ldots, T\}$ that covers bps $i$ to $j$ on the lattice be $u_t(i,j)$.
The interaction between a particle of type $t$ ending at position $k$, and the next particle of type $s$ starting at position $l$
will be denoted by $\Phi(t,k;s,l)$ (Fig. \ref{Fig:NucsTF}). Each particle of type $t$, when attached to the DNA, is in contact with a number
of bps between $a^t_\text{min}$ and $a^t_\text{max}$. Thus $u_t(i,j) = 0$ if $i$ and $j$ do not satisfy the constraints
$a^t_\text{min}\leq j-i+1 \leq a^t_\text{max}$. Also, $\Phi(t,i;s,j)=\infty$ for $j\leq i$ since the particles cannot overlap.
With this notation, the grand-canonical partition function becomes
\[
	Z = \sum_{\text{all states}} e^{-\beta [u_{t1}(i_{1L},i_{1R})-\mu_{t_1}]} e^{-\beta \Phi(t_1,i_{1R};t_2,i_{2L})} e^{-\beta [u_{t_2}(i_{2L},i_{2R})-\mu_{t_2}]} \ldots,
\]
where $\mu_t$ is the chemical potential of the particles of type $t$. The sum is over all configurations, which can have variable numbers of particles of any type.

Using the matrix notation,
\begin{align*}
	\la t,k \ve z \ve s,l\ra &= e^{-\beta [u_t(k,l)-\mu_t]} \delta_{t,s}
	\equiv \zeta_t(k,l) \delta_{t,s},\\
	\la t,k \ve w \ve s,l\ra &= e^{-\beta \Phi(t,k;s,l)},
\end{align*}
where $\delta_{t,s}$ is the Kronecker delta symbol, the partition function can be written as
\[
Z = \sum_{\text{all states}} \la t_1, i_{1L} \ve z \ve t_1,i_{1R} \ra \la t_1,i_{1R} \ve w \ve t_2,i_{2L}  \ra \la t_2,i_{2L} \ve z \ve t_2,i_{2R} \ra \ldots
\]

Each vector $\ve t,i \ra$ has dimension $T L$ and only one non-zero element, set to 1 for normalization.
For example, $\ve 1,i \ra$ vectors have a 1 at position $i$, $\ve 2,i \ra$ vectors have a 1 at position $L+i$, etc.
As above, we denote by $\ve J \ra$ a vector in which all $T L$ elements are equal to 1, to obtain
\[
Z = 1 + \la J \ve (I-z w)^{-1} z \ve J \ra,
\]
equivalent to Eq.~\eqref{Eq:Z}.

Similarly to the previous case of a single particle type, we can compute the one-particle density
\begin{eqnarray} \label{Eq:nMultiTypes}
n^t_1(k,l) &=& \frac{\zeta_t(k,l)}{Z} \frac{\delta Z}{\delta \zeta_t(k,l)} \nonumber \\
                           &=& \frac{1}{Z} \la J \ve (I - z w)^{-1} \ve t,k \ra \la t,k \ve z \ve t,l \ra \la t,l \ve (I - w z)^{-1} \ve J \ra.
\end{eqnarray}

We can also obtain the two-particle density
\begin{align} \label{Eq:n2MultiTypes}
n^{t,s}_2(i,j;k,l) &= \frac{1}{Z} \la J \ve (I - z w)^{-1} \ve t,i \ra \la t,i \ve z \ve t,j \ra \nonumber \\
				&\quad \times \la t,j \ve w (I - z w)^{-1} \ve s,k \ra \la s,k \ve z \ve s,l \ra \la s,l \ve (I- w z)^{-1} \ve J \ra
\end{align}
and the nearest-neighbor two-particle density
\begin{align} \label{Eq:n2MultiTypes:nn}
	\overline{n}^{t,s}_2(i,j;k,l) &= \frac{1}{Z} \la J \ve (I - z w)^{-1} \ve t,i \ra \la t,i \ve z \ve t,j \ra \nonumber \\
				&\quad \times \la t,j \ve w \ve s,k \ra \la s,k \ve z \ve s,l \ra \la s,l \ve (I- w z)^{-1} \ve J \ra.
\end{align}
Eqs.~\eqref{Eq:n2MultiTypes} and \eqref{Eq:n2MultiTypes:nn} give the joint probability that a particle of type $t$ covers bps $i$ to $j$, while a second particle of type $s$ covers bps $k$ to $l$.

Using Eq.~\eqref{Eq:nMultiTypes} we can compute occupancy for each type of particles $t$ and for each bp $i$:
\[
\operatorname{Occ}_t (i) = \sum_{k=i-a^t_\text{max}+1}^i \sum_{l=\max(i, k+a^t_\text{min})}^{k+a^t_\text{max}-1} n^t_1(k,l).
\]

In the following sections we will focus on the one-particle density function.

\subsection{Direct problem: Recursive solution for hard-core interactions} \label{section:DP_RecForm}

A straightforward application of Eqs.~\eqref{Eq:nMultiTypes} and \eqref{Eq:n2MultiTypes} entails computationally intensive
matrix manipulations. Fortunately, for particles that interact only through hard-core repulsion rather than long-range two-body interactions, the one-particle distribution can be computed recursively and therefore much more efficiently.

\subsubsection{General case} \label{section:DP_RecForm_1Sp}
With multiple particle types, Eq.~\eqref{Eq:nMultiTypes} can be rewritten as
\[
n^t_1(i,j) = \frac{1}{Z} Z^{-}(i) \la t,i \ve z \ve t,j \ra Z^{+}(j),
\]
where $Z^{-}(i)$ and $Z^{+}(j)$ are the partition functions for the domains $[1, i)$ and
$(j, L]$, respectively. Note that in the case of hard-core interactions alone, $Z^{-}(i)$ and $Z^{+}(j)$ do not depend on the type of the particle occupying positions $i$ through $j$.

In the case of steric exclusion alone, these partial partition functions satisfy the following recursion relations:
\begin{equation}
Z^{-}(i) = Z^{-}(i-1) + \sum_s \sum_{i-a^s_\text{max}\leq j\leq i-a^s_\text{min}} Z^{-}(j) \la s,j \ve z \ve s,i-1 \ra, \label{Eq:Rec12}
\end{equation}
and
\begin{equation}
Z^{+}(i) = Z^{+}(i+1) + \sum_s \sum_{i+a^s_\text{min}\leq j\leq i+a^s_\text{max}} \la s,i+1 \ve z \ve s,j \ra Z^{+}(j). \label{Eq:Rec22}
\end{equation}
Here each particle type $s$ has two characteristic lengths, corresponding to its minimum and maximum DNA footprints, respectively:
$a^s_\text{min}$ and $a^s_\text{max}$.
The boundary conditions are $Z^-(1)=1$ and $Z^+(L)=1$. The full partition function is given by $Z = Z^-(L+1) = Z^+(0)$. Note that all unphysical terms
for which bound particles run off the lattice automatically vanish from Eqs.~\eqref{Eq:Rec12} and \eqref{Eq:Rec22}.
To avoid numeric instabilities, the recursion should be done in log space. Let

\begin{align*}
F(i) &= \ln Z^-(i),\\
R(i) &= \ln Z^+(i).
\end{align*}

With this notation, Eqs.~\eqref{Eq:Rec12} and \eqref{Eq:Rec22} become
\begin{align} \label{Eq:Rec32}
F(i) &= F(i-1) + \ln \Big\{ 1
 + \sum_s \sum_{i-a^s_\text{max}\leq j\leq i-a^s_\text{min}} e^{F(j) - F(i-1) + \beta\left[ \mu_s - u_s(j, i-1)\right]}\Big\}, \\ \nonumber
R(i) &= R(i+1) + \ln \Big\{ 1
 + \sum_s \sum_{i+a^s_\text{min}\leq j\leq i+a^s_\text{max}} e^{R(j) - R(i+1) + \beta\left[ \mu_s - u_s(i+1, j)\right]}\Big\},
\end{align}
with the boundary conditions $F(1) = R(L) = 0$.

Then the one-particle distribution function is
\[
n^t_1 (i,j) = e^{F(i) + R(j) - \ln Z + \beta \left[ \mu_t - u_t(i,j) \right]},
\]
where $\ln Z = F(L+1) = R(0)$.

Although we do not show it here, the two-particle distribution can be computed in a very similar way.
The only new ingredient in Eq.~\eqref{Eq:n2MultiTypes} is the partition function for the box with walls at two arbitrary positions,
$Z(t,j,s,k) \equiv \la t,j \ve w (I - z w)^{-1} \ve s,k \ra$. This partition function can be computed recursively,
exactly as the partial partition functions $Z^\pm$ discussed above.

\subsubsection{Special case: No unwrapping}

The special case in which all particles are fully attached to their DNA sites (i.e., there is no DNA unwrapping) can be easily obtained from our general formalism.
Indeed, in this case we restrict $a^s_\text{min} = a^s_\text{max} = a^s$ in Eq.~\eqref{Eq:Rec32}, obtaining

\begin{align*}
F(i) &= F(i-1) + \ln \Big\{ 1 + \sum_s e^{F(i-a^s) - F(i-1) + \beta\left[ \mu_s - u_s(i-a^s, i-1)\right]}\Big\},\\
R(i) &= R(i+1) + \ln \Big\{ 1 + \sum_s e^{R(i+a^s) - R(i+1) + \beta\left[ \mu_s - u_s(i+1, i+a^s)\right]}\Big\}.
\end{align*}
As before, the boundary conditions are $F(1) = R(L) = 0$.

The one-particle distribution is given by
\[
n^t_1 (i,i+a^t-1) = e^{F(i) + R(i+a^t-1) - \ln Z + \beta \left[ \mu_t - u_t(i,i+a^t-1) \right]}.
\]

\subsection{Inverse problem: Recursive solution for hard-core interactions}

In the previous section, we have solved the direct problem: given the binding energies for all particle types,
we compute s-particle distributions. However, typically it is particle distributions that are observed experimentally, and the
energetics of particle-DNA interactions need to be inferred. This inverse problem can be solved recursively for the case
of systems with multiple particle types, partial unwrapping (variable footprints), and steric exclusion. The recursive solution
is efficient enough to be employed on the genome-wide scale. Here we focus on one-particle distributions and one-body 
energies; the exact matrix formulation of the inverse problem for a single particle type with the two-particle distribution and the two-body potential is available
in Ref.~\cite{Chereji2011b}.

\subsubsection{General case} \label{subsection:Theory42}

Using Eqs.~\eqref{Eq:nMultiTypes}, \eqref{Eq:Rec12} and \eqref{Eq:Rec22}, we obtain:

\begin{align}
Z^-(i) &= Z^-(i-1) \left[1 + \sum_{\substack{t,\\i-a^t_\text{max}\leq j\leq i-a^t_\text{min}}} \frac{Z}{Z^-(i-1) Z^+(i-1)} n^t_1(j,i-1) \right] \notag\\
           &= Z^-(i-1) \left[1 + \frac{N^R(i-1)}{\xi(i-1)}\right], \label{Eq:Rec7}
\end{align}
where $N^R(i) = \sum_t \sum_{i-a^t_\text{max}+1\leq j\leq i-a^t_\text{min}+1} n^t_1 (j,i)$
represents the probability of finding a particle of any type with the right edge at bp $i$, and $\xi(i)=Z^-(i) Z^+(i)/Z$.

$Z^+$ satisfies a similar recursive relation:
\begin{equation}
Z^+(i) = Z^+(i+1) \left[ 1 + \frac{N^L(i+1)}{\xi(i+1)}\right], \label{Eq:Rec8}
\end{equation}
where $N^L(i)$ is the probability of finding a particle of any type with the left edge at bp $i$:
$N^L(i) = \sum_t \sum_{i+a^t_\text{min}-1\leq j\leq i+a^t_\text{max}-1}
n^t_1 (i,j)$.

The quantity $\xi(i)$ satisfies
\begin{align*}
\xi(i+1) - \xi(i) =& \frac{1}{Z} \left[ Z^-(i+1)Z^+(i+1) - Z^-(i)Z^+(i) \right] \\
	          =& \frac{1}{Z} \Big\{ Z^-(i+1)\left[Z^+(i+1) - Z^+(i)\right] \\
	             & + Z^+(i) \left[ Z^-(i+1) - Z^-(i) \right]\Big\} \\
	          =& N^R(i) - N^L(i+1),
\end{align*}

so that
\begin{equation} \label{xi:recurs}
\xi(i) = 1 + \sum_{k=0}^{i-1} \left[ N^R(k) - N^L(k+1)\right],
\end{equation}
where the initial condition $\xi(0)=1$ has been used.

After we compute both $Z^-$ and $Z^+$ in this way, the total partition function is given by $Z = Z^-(L+1) = Z^+(0)$ as before,
and the binding energy for any particle attached to the DNA is given by
\begin{equation}
\beta \left[ u_t(i,j) - \mu_t \right] = - \ln \left[ n^t_1 (i,j) \frac{Z}{Z^{-}(i) Z^{+}(j)} \right]. \label{Eq:u2}
\end{equation}

\subsubsection{Special case: No unwrapping} \label{subsection:Theory52}

In the case of the all-or-none binding, all matrix elements $\la i \ve n^t_1 \ve j \ra$ vanish unless $j = i + a^t - 1$,
where $a^t$ is the length of the binding site for the particle of type $t$. Thus
\begin{align*}
N^L(i) &= \sum_t n^t_1 (i, i + a^t - 1),\\
N^R(i) &= \sum_t n^t_1 (i - a^t + 1, i).
\end{align*}

Using these expressions, we can employ Eqs.~\eqref{Eq:Rec7}, \eqref{Eq:Rec8} and \eqref{xi:recurs} to compute $Z^+$ and $Z^-$ in log space.
Finally, Eq.~\eqref{Eq:u2} can be used to compute the binding energies.

If all particles are of the same type, the quantity $\xi$ can be simplified further:
\[
\xi(i) = 1 - \sum_{k=i-a+1}^{i} N^L(k) = 1 - \operatorname{Occ}(i),
\]
where $\operatorname{Occ}(i)$ is the probability that bp $i$ is covered by a particle. Thus in this limit $\xi (i)$ is simply the probability that
bp $i$ is not bound by any particles.

The recursion relations for $Z^-$ and $Z^+$ become
\begin{align*}
&Z^-(i+1) = Z^-(i) \left[ 1 + \frac{N^L(i - a + 1)}{1 - \operatorname{Occ}(i)} \right], \\
&Z^+(i) = Z^+(i+1) \left[ 1 + \frac{N^L(i+1)}{1 - \operatorname{Occ}(i + 1)} \right].
\end{align*}
These expressions are equivalent to those previously obtained in \cite{Locke2010}.

\subsection{Sequence-specific nucleosome formation energies} \label{subsection:Theory6}

The simplest sequence-dependent model of nucleosome formation assumes that the nucleosome energy depends only on the mono- and dinucleotide counts
in the nucleosomal site~\cite{Locke2010}. Thus for the sequence $S_1 S_2\ldots S_N$, the nucleosome formation energy is
$\sum_{i=1}^N \epsilon_{S_i} + \sum_{i=1}^{N-1} \epsilon_{S_i S_{i+1}}$, where $\epsilon_{S_i}$ and $\epsilon_{S_i S_{i+1}}$ are
the mono- and dinucleotide contributions, respectively.

Here we consider a symmetrized version of the model, with $\epsilon_{A} = \epsilon_{T}$, $\epsilon_{C} = \epsilon_{G}$ and 10
unique dinucleotide energies:  $\epsilon_{AA/TT}$, $\epsilon_{AC/GT}$, $\epsilon_{AG/CT}$, $\epsilon_{AT}$, $\epsilon_{CA/TG}$, $\epsilon_{CC/GG}$, $\epsilon_{CG}$, $\epsilon_{GA/TC}$, $\epsilon_{GC}$, $\epsilon_{TA}$.

Using Eq.~\eqref{Eq:u2}, we can estimate all 12 $\epsilon$'s for a model in which nucleosomes are the only particle type and unwrapping is allowed.
For each sequence starting a bp $i$ and ending at bp $j$, we have
\begin{align}
u(i,j) - \mu &= \sum_{k=i}^j \epsilon_{S_k} + \sum_{k=i}^{j-1} \epsilon_{S_k S_{k+1}} - \mu,\notag\\
&= 
\left(\begin{array}{c c c c c c}
m_{A/T} & m_{C/G} & m_{AA/TT} & \cdots & m_{TA} & -1
\end{array}\right)
\left(
\begin{array}{c}
	\epsilon_{A/T} \\
	\vdots \\
	\epsilon_{TA} \\
	\mu
\end{array}
\right),\label{Eq:EnergyForSequence}
\end{align}
where $m_X$ and $m_{XY}$ are the counts of mono- and dinucleotides $X$ and $XY$ in the sequence, respectively.

Using all possible combinations of pairs $(i,j)$ where a nucleosome can form, we obtain a large number $P$ of equations of this type:
\[
\begin{array}{c}
	 E - \mu 
\end{array}
=
\begin{array}{c}
	M 
\end{array}
\left(
\begin{array}{c}
	\epsilon \\
	\mu 
\end{array}
\right).
\]
Here, $E - \mu$ is a column vector of dimension $P$, where each row contains one $u(i,j) - \mu$ element from Eq.~\eqref{Eq:EnergyForSequence}.
$\left(
\begin{array}{c}
	\epsilon \\
	\mu 
\end{array}
\right)$
is the column vector from Eq.~\eqref{Eq:EnergyForSequence}, and $M$ is a $P \times 13$ matrix with mono- and dinucleotide counts and -1's in the last column.
From this equation we can derive the energies $\epsilon$ and $\mu$ by a least squares fit.

A problem that arises is that matrix $M$ has a one-dimensional kernel, or null space~\cite{Chereji2011b}.
Because in every DNA sequence the number of mononucleotides is equal to the length of the sequence,
and the number of dinucleotides is equal to the length of the sequence minus 1, the columns of the matrix $M$ are not linearly independent.
Indeed, the column vector
\[
\ve V \ra =
\left(
\begin{array}{c}
1 \\
1 \\
-1 \\
\vdots \\
-1 \\
1
\end{array}
\right)
\]
is the only linearly independent vector from the kernel of $M$: $M \ve V \ra = 0$
(i.e., the kernel of $M$ is spanned by $\ve V \ra$).
Thus we have only 12 linearly independent equations, and cannot obtain a unique solution for the 13 parameters $\epsilon$ and $\mu$.
We need to add one constraint, which we choose to be
 \[
\sum_{i=1}^{12} \epsilon_i = 0,
\]
to make the solution unique.

\subsection{Site-specific chemical cleavage bias} \label{subsection:CleavageFilter}

Hydroxyl radicals that cleave DNA near the nucleosome dyad have two preferred cutting sites,
at positions -1 bp and +6 bp with respect to the dyad \cite{Brogaard2012}. If DNA is cut at these positions with frequencies $f$ and $1-f$, the distance between two consecutive cuts (one on the Watson and one on the Crick strand) is given by
\[
d_\text{cuts} = d_\text{dyads} + b,
\]
where $d_\text{dyads}$ is the distance between two neighboring dyads and the bias $b$ is
\[
b = 
\left\{
\begin{array}{l}
-12\\
-5\\
2
\end{array}\right.
\textrm{ with probability }
\left\{
\begin{array}{l}
(1-f)^2\\
2f(1-f)\\
f^2
\end{array}\right. .
\]

The cleavage bias has to be taken into account by convolving the predicted inter-dyad distance probability $P(c+d|c)$ 
with a kernel corresponding to this bias:
\[
F(x) = 
\left\{
\begin{array}{l l}
(1-f)^2 & \textrm{for }x=-12\\
2f(1-f) & \textrm{for }x=-5\\
f^2     & \textrm{for }x=2\\
0       & \textrm{otherwise}
\end{array}\right. .
\]

The convolution is then compared with the observed distribution of inter-dyad distances.

\subsection{Parameter optimization} \label{subsection:ParamOpt}

Parameter fitting for all models was carried out in a two-stage procedure using the genetic algorithm optimization function \texttt{ga}
from the MATLAB Global optimization toolbox. First, the objective function to be minimized was set equal to the root-mean-square (RMS) deviation
between predicted and observed inter-dyad distributions. Once $RMS < 10^{-3}$ had been achieved, the objective function was replaced by
$\text{RMS} - r_{\text{osc}} \simeq - r_{\text{osc}}$,
where $r_{\text{osc}}$ is the linear correlation between observed and predicted oscillations after the smooth background has been subtracted from
inter-dyad distributions, as in Fig.~1D. We have found that the two-stage setup allows us to effectively fit both the overall shape and the fine oscillatory structure in the data.
The best-fit parameters for all models are given in SI Results.

\newpage
\section{Supplementary Results}

\renewcommand{\thetable}{S\arabic{table}}
\setcounter{table}{0}

{\bf Model A: Crystal structure augmented with an additional well.} The binding energy of a particle of length $a = 1 + x_1 + x_2$ (1 bp for the dyad, and $x_1$ and $x_2$ for the extra number of bps in contact with the histone octamer on each side of the dyad) is given by $u = u_\textrm{half}(x_1) + u_\textrm{half}(x_2)$, with
\[
u_\textrm{half}(x) = \texttt{interp1}(...) - \frac{E_b}{147} x,
\]
where $E_b$ is the binding energy of a fully wrapped particle in the absence of 10-11 bp oscillations. The oscillations are based on the the positions of the histone-DNA contacts in the crystal structure~\cite{Richmond2003}. The MATLAB function \texttt{interp1}(...) was used to generate the oscillatory pattern by piecewise cubic Hermite interpolation between the following data points:
\begin{table}[h!]
\centering
\begin{dataTable}%
{@{\hspace{2ex}} c @{\hspace{6ex}} c @{\hspace{2ex}}}%
{\scshape\textcolor{white}{x (Position)}} &
{\scshape\textcolor{white}{f(x) (Energy)}}\\ \midrule[0pt]
  -1 & -A\\ \midrule
   3 &  A\\ \midrule
   7 & -A\\ \midrule
  13 &  A\\ \midrule
  17 & -A\\ \midrule
  24 &  A\\ \midrule
  28 & -A\\ \midrule
  34 &  A\\ \midrule
  38 & -A\\ \midrule
  44 &  A\\ \midrule
  49 & -A\\ \midrule
  55 &  A\\ \midrule
  59 & -A\\ \midrule
  65 &  A\\ \midrule
  69 & -A\\ \midrule
  75 &  A\\ \midrule
   p & -d\\ \midrule
  85 &  A
\end{dataTable}
\end{table}

\noindent
The oscillatory pattern was superimposed onto a line with the slope of
$-E_b/147$.

\newpage
\begin{table}[h!]
\begin{minipage}[t]{.5\textwidth}
\vspace{3.5pt}
\centering
\begin{dataTable}%
{@{\hspace{2ex}} l @{\hspace{6ex}} l @{\hspace{2ex}}}%
{\scshape\textcolor{white}{Parameter}} &
{\scshape\textcolor{white}{Value}}\\ \midrule[0pt]
  $a_\textrm{max}$ & 163 bp\\ \midrule
  $a_\textrm{min}$ & 3 bp\\ \midrule
  $E_b$ & 14.39 $\textrm{k}_\textrm{B}\textrm{T}$\\ \midrule
  $\mu$ & -14.51 $\textrm{k}_\textrm{B}\textrm{T}$\\ \midrule
  $A$   & 1.13 $\textrm{k}_\textrm{B}\textrm{T}$\\ \midrule
  $f$ & 0.51\\ \midrule
  $p$ & 79 bp\\ \midrule
  $d$ & 0.86 $\textrm{k}_\textrm{B}\textrm{T}$
\end{dataTable}
\end{minipage}
\hfill
\begin{minipage}[t]{.45\textwidth}
\vspace{0pt}
\centering
\includegraphics[width=\textwidth]{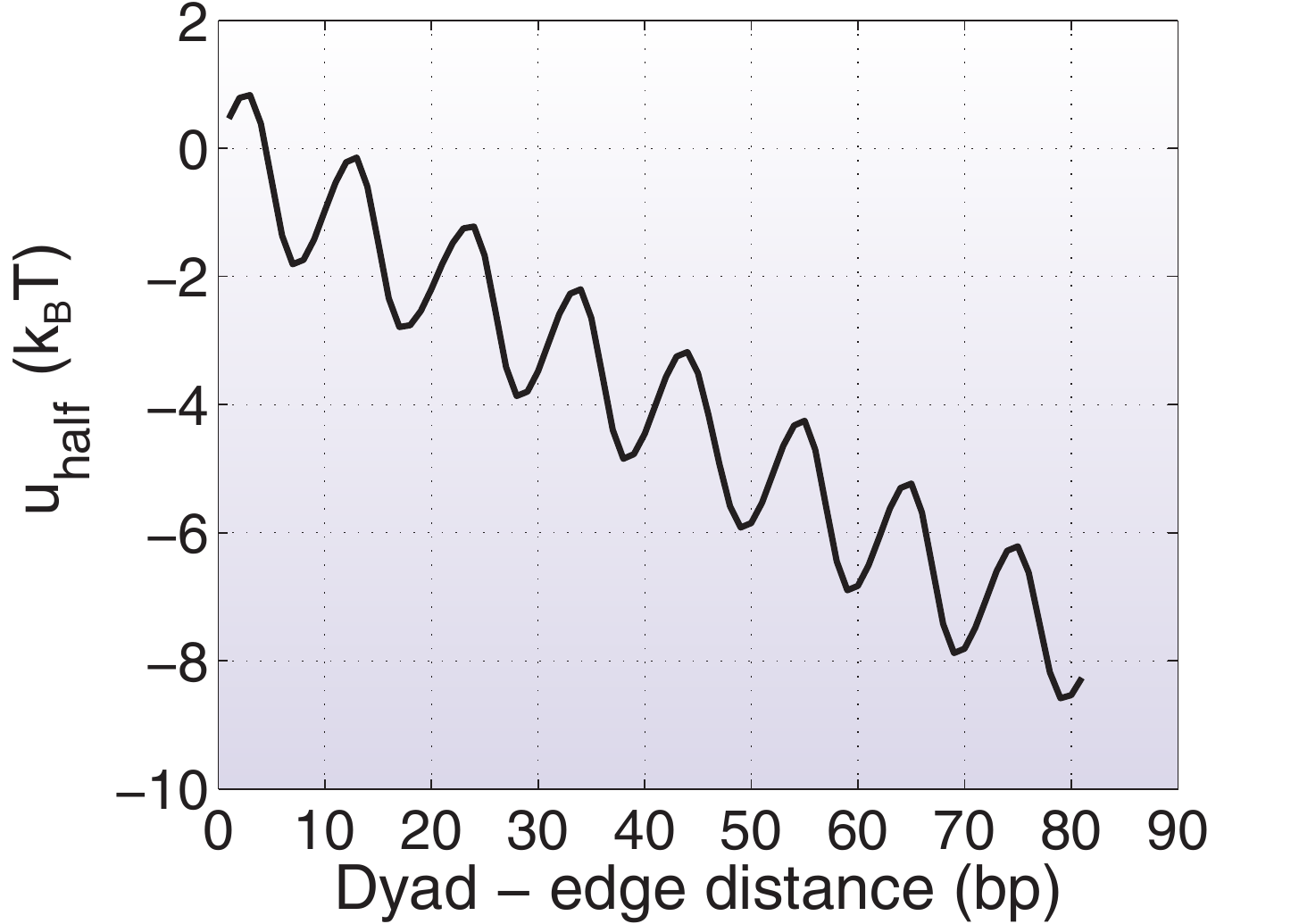}
\end{minipage}
\hspace*{\fill}
\caption{Fitted parameters for Model A. $a_\textrm{max}$ and $a_\textrm{min}$ are the maximum and minimum lengths of the nucleosome particle, $\mu$ is the histone octamer chemical potential, $A$ is the amplitude of the oscillations, and $f$ is the hydroxyl radical cutting frequency. $p$ and $d$ are the position and the depth of the first minimum outside of the nucleosome core particle, respectively. $E_b$ is the binding energy of a fully wrapped particle in the absence of 10-11 bp oscillations.}
\end{table}

\vspace{1cm}
Fit residuals:
\begin{align*}
&RMS = 9.9958 \times 10^{-4}\\
&r_{\text{osc}} = 0.764364\\
&RMS_{\text{osc}} = 1.8662 \times 10^{-4}
\end{align*}
$RMS$ is the root-mean-square error of the predicted inter-dyad distribution.
$r_{\text{osc}}$ is the linear correlation between the oscillatory parts of the measured and predicted inter-dyad distributions. The oscillatory part was obtained by subtracting the smooth background from the full inter-dyad distribution. Smoothing was done by applying a Savitzky-Golay smoothing filter [also known as least-squares, or DISPO (Digital Smoothing Polynomial) filter] of polynomial order 3 and length 31 bp. $RMS_{\text{osc}}$ is the root-mean-square error of the oscillatory part of the predicted inter-dyad distribution.\\

\newpage

{\bf Model B: Crystal structure augmented with a linear function.} Same as Model A for $x \in [1,73]$, followed by a linear function:
\[
u_\textrm{half}(x) = 
\left\{
\begin{array}{ll}
\texttt{interp1}(...) - \frac{E_b}{147} x &\text{for } x \in [1, 73],\\ 
\texttt{interp1}(...) - 73 \frac{E_b}{147} - \frac{\Delta E}{\Delta X} (x - 73) &\text{for } x \in [74, 73 + \Delta X],
\end{array}
\right.
\]
where ${\Delta E}/{\Delta X}$ is the slope of the linear function (i.e., ${\Delta E}$ is the energy difference between the first and last points of the linear function and ${\Delta X}$ is the cardinality of the range of the linear function).

\begin{table}[h!]
\begin{minipage}[t]{.5\textwidth}
\vspace{3.5pt}
\centering
\begin{dataTable}%
{@{\hspace{2ex}} l @{\hspace{6ex}} l @{\hspace{2ex}}}%
{\scshape\textcolor{white}{Parameter}} &
{\scshape\textcolor{white}{Value}}\\ \midrule[0pt]
  $a_\textrm{min}$ & 27 bp\\ \midrule
  $E_b$ & 14.66 $\textrm{k}_\textrm{B}\textrm{T}$\\ \midrule
  $\mu$ & -15.04 $\textrm{k}_\textrm{B}\textrm{T}$\\ \midrule
  $A$   & 1.28 $\textrm{k}_\textrm{B}\textrm{T}$\\ \midrule
  $f$ & 0.50\\ \midrule
  $\Delta E$ & -2.47 $\textrm{k}_\textrm{B}\textrm{T}$\\ \midrule
  $\Delta X$ & 7 bp
\end{dataTable}
\end{minipage}
\hfill
\begin{minipage}[t]{.45\textwidth}
\vspace{0pt}
\centering
\includegraphics[width=\textwidth]{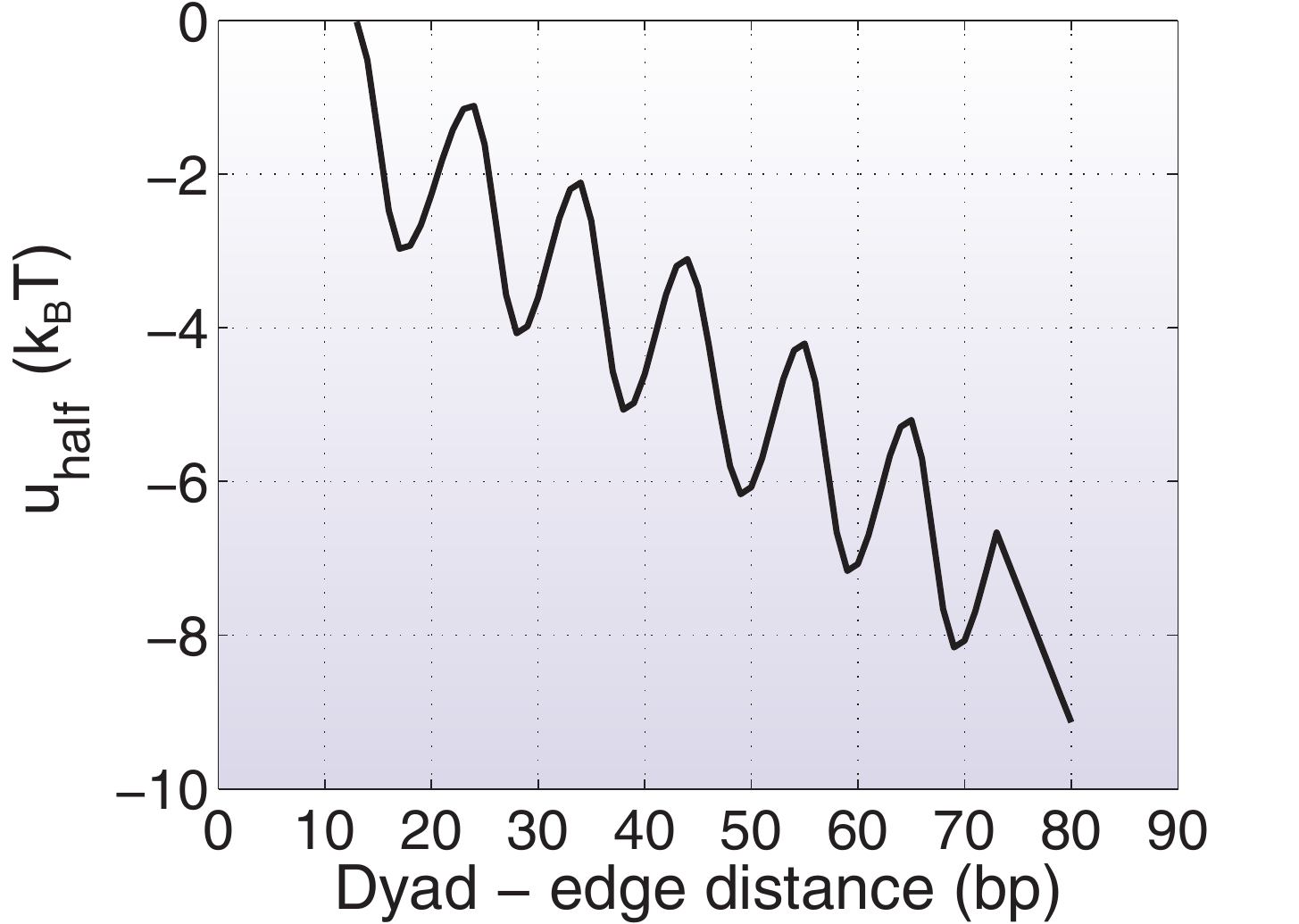}
\end{minipage}
\hspace*{\fill}
\caption{Fitted parameters for model B. All parameters are as in Model A, except for ${\Delta E}$ and ${\Delta X}$ which are defined above.}
\end{table}

Fit residuals:
\begin{align*}
&RMS = 0.0012 \textrm{ ($RMS$ cannot decrease below $10^{-3}$ for this model)}\\
&r_{\text{osc}} = 0.769179\\
&RMS_{\text{osc}} = 1.8411 \times 10^{-4}
\end{align*}
All residuals are defined as in Model A.

\newpage
{\bf Model C: 10-bp oscillations superimposed onto a linear function.}
\[
u_\textrm{half}(x) = - A \cos\left(\frac{2 \pi}{10} (x - x_0)\right) - \frac{E_b}{147} x,
\]
where $A$ is the amplitude of the oscillations, $x_0$ determines the phase of the oscillations, and $E_b$ is the binding energy of a fully wrapped particle in the absence of the oscillations.

\begin{table}[h!]
\begin{minipage}[t]{.5\textwidth}
\vspace{3.5pt}
\centering
\begin{dataTable}%
{@{\hspace{2ex}} l @{\hspace{6ex}} l @{\hspace{2ex}}}%
{\scshape\textcolor{white}{Parameter}} &
{\scshape\textcolor{white}{Value}}\\ \midrule[0pt]
  $a_\textrm{max}$ 	& 165 bp\\ \midrule
  $a_\textrm{min}$ 	& 3 bp\\ \midrule
  $E_b$ 			& 14.43 $\textrm{k}_\textrm{B}\textrm{T}$\\ \midrule
  $\mu$ 			& -13.99 $\textrm{k}_\textrm{B}\textrm{T}$\\ \midrule
  $A$ 				& 1.06 $\textrm{k}_\textrm{B}\textrm{T}$\\ \midrule
  $x_0$ 			& 79 bp\\ \midrule
  $f$ 				& 0.50
\end{dataTable}
\end{minipage}
\hfill
\begin{minipage}[t]{.45\textwidth}
\vspace{0pt}
\centering
\includegraphics[width=\textwidth]{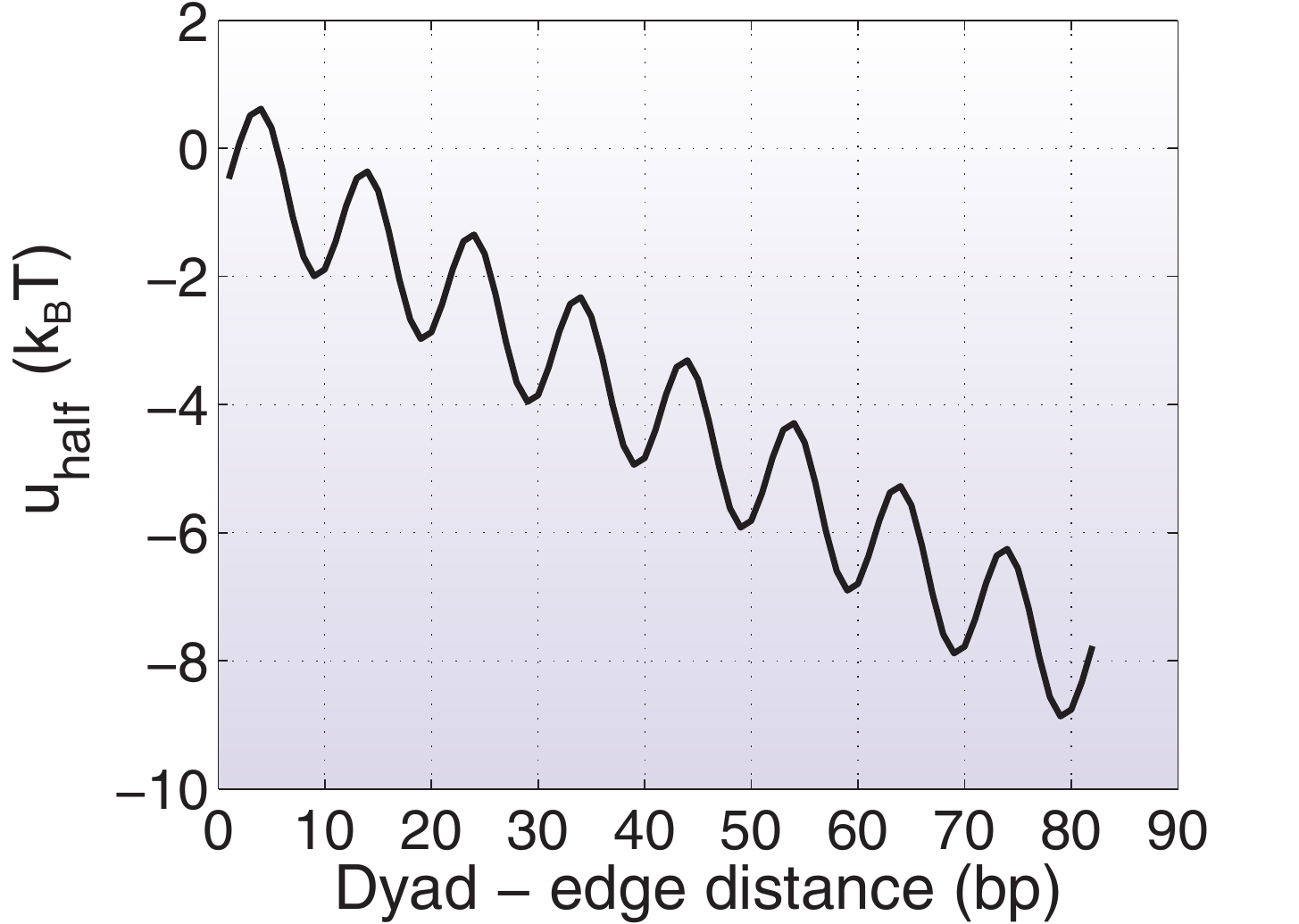}
\end{minipage}
\hspace*{\fill}
\caption{Fitted parameters for Model C. All parameters are as in Model A, except for $x_0$ which is defined above.}
\end{table}

Fit residuals:
\begin{align*}
&RMS = 9.9861 \times 10^{-4}\\
&r_{\text{osc}} = 0.708620\\
&RMS_{\text{osc}} = 2.0203 \times 10^{-4}
\end{align*}
All residuals are defined as in Model A.

\newpage
{\bf Model D: 11-bp oscillations superimposed onto a linear function.}
\[
u_\textrm{half}(x) = - A \cos\left(\frac{2 \pi}{11} (x - x_0)\right) - \frac{E_b}{147} x,
\]
where $A$, $x_0$ and $E_b$ have the same meaning as in Model C.

\begin{table}[h!]
\begin{minipage}[t]{.5\textwidth}
\vspace{3.5pt}
\centering
\begin{dataTable}%
{@{\hspace{2ex}} l @{\hspace{6ex}} l @{\hspace{2ex}}}%
{\scshape\textcolor{white}{Parameter}} &
{\scshape\textcolor{white}{Value}}\\ \midrule[0pt]
  $a_\textrm{max}$ 	& 161 bp\\ \midrule
  $a_\textrm{min}$ 	& 25 bp\\ \midrule
  $E_b$ 			& 13.99 $\textrm{k}_\textrm{B}\textrm{T}$\\ \midrule
  $\mu$ 			& -14.30 $\textrm{k}_\textrm{B}\textrm{T}$\\ \midrule
  $A$ 				& 1.03 $\textrm{k}_\textrm{B}\textrm{T}$\\ \midrule
  $x_0$			 	& 80 bp\\ \midrule
  $f$ 				& 0.52
\end{dataTable}
\end{minipage}
\hfill
\begin{minipage}[t]{.45\textwidth}
\vspace{0pt}
\centering
\includegraphics[width=\textwidth]{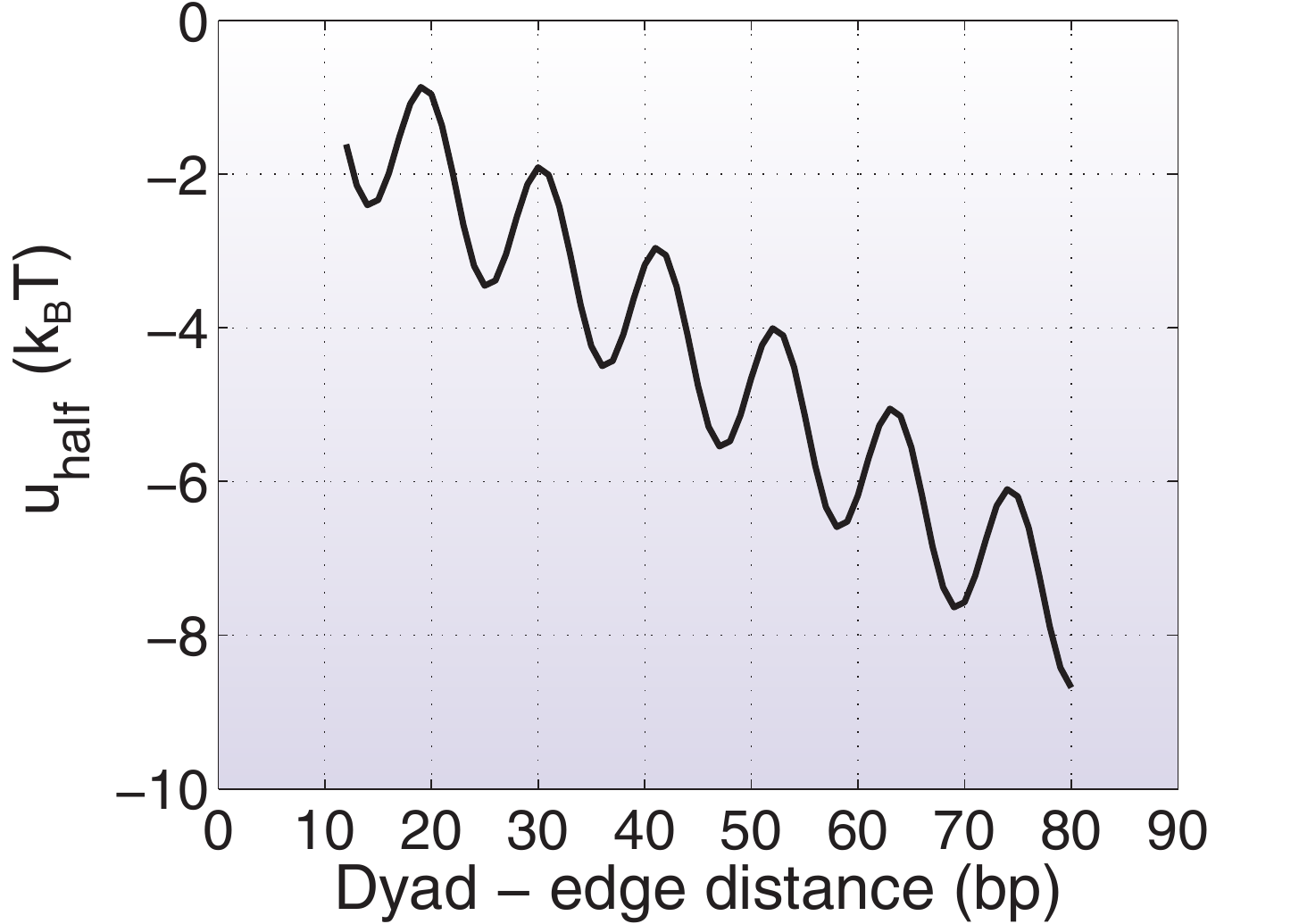}
\end{minipage}
\hspace*{\fill}
\caption{Fitted parameters for Model D. All parameters are as in Model C.}
\end{table}

Fit residuals:
\begin{align*}
&RMS = 9.9121 \times 10^{-4}\\
&r_{\text{osc}} = 0.688838\\
&RMS_{\text{osc}} = 2.0735 \times 10^{-4}
\end{align*}
All residuals are defined as in Model A.

\newpage
{\bf Model E: Uniform unwrapping.}
\[
u_\textrm{half}(x) = - \frac{E_b}{147} x,
\]
where $E_b$ is the binding energy of a fully wrapped nucleosome.

\begin{table}[h!]
\begin{minipage}[t]{.5\textwidth}
\vspace{3.5pt}
\centering
\begin{dataTable}%
{@{\hspace{2ex}} l @{\hspace{6ex}} l @{\hspace{2ex}}}%
{\scshape\textcolor{white}{Parameter}} &
{\scshape\textcolor{white}{Value}}\\ \midrule[0pt]
  $a_\textrm{max}$ & 163 bp\\ \midrule
  $a_\textrm{min}$ & 35 bp\\ \midrule
  $E_b$ & 13.40 $\textrm{k}_\textrm{B}\textrm{T}$\\ \midrule
  $\mu$ & -13.14 $\textrm{k}_\textrm{B}\textrm{T}$\\ \midrule
  $f$ & 0.58
\end{dataTable}
\end{minipage}
\hfill
\begin{minipage}[t]{.45\textwidth}
\vspace{0pt}
\centering
\includegraphics[width=\textwidth]{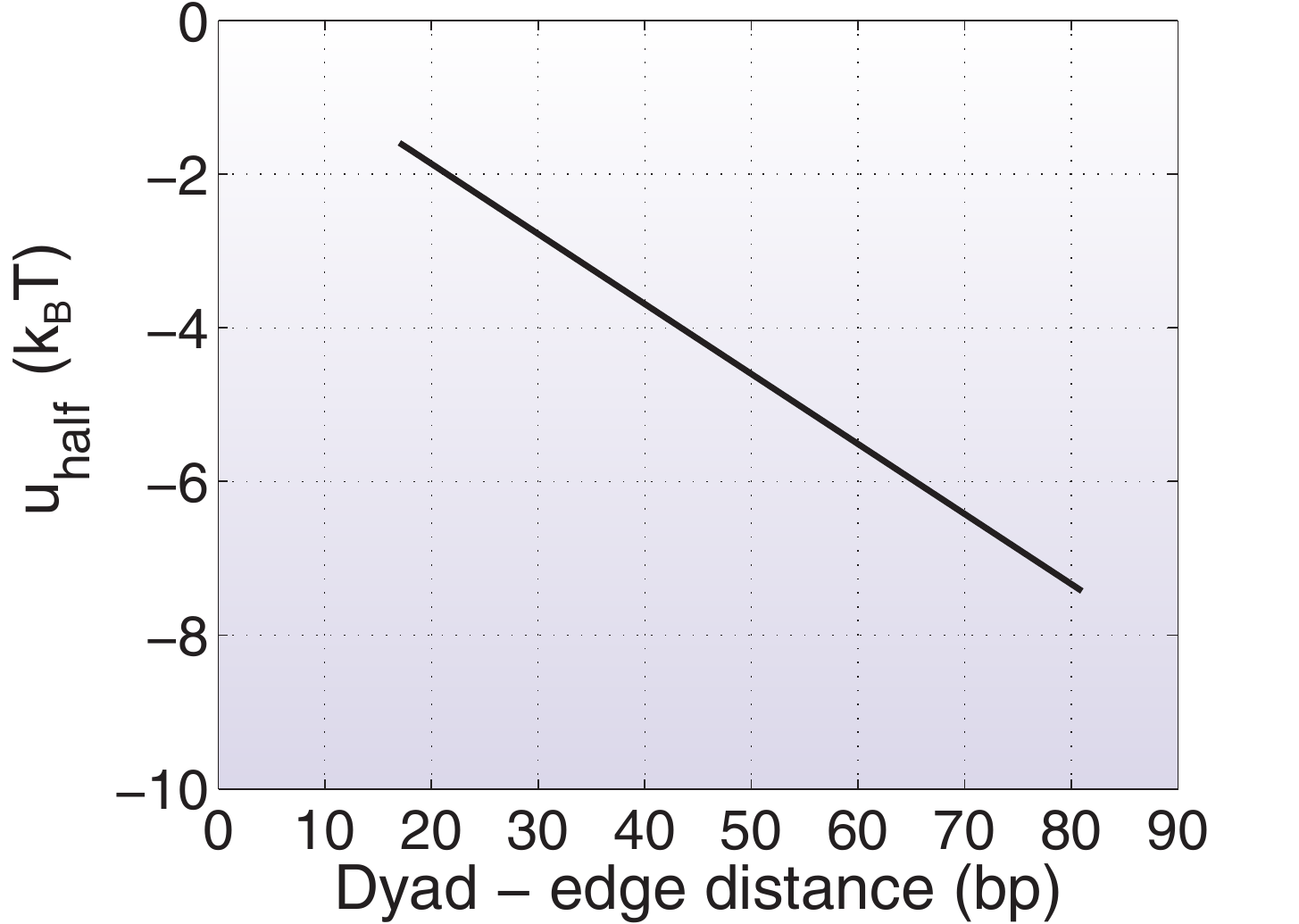}
\end{minipage}
\hspace*{\fill}
\caption{Fitted parameters for Model E. All parameters are as in Model A.}
\end{table}

Fit residuals:
\begin{align*}
&RMS = 9.9725 \times 10^{-4}\\
&r_{\text{osc}} = 0.274786\\
&RMS_{\text{osc}} = 2.7507 \times 10^{-4}
\end{align*}
All residuals are defined as in Model A.

\newpage
{\bf Model F: 5-bp oscillations superimposed onto a linear function.}
\[
u_\textrm{half}(x) = - A \cos\left(\frac{2 \pi}{5} (x - x_0)\right) - \frac{E_b}{147} x,
\]
where $A$, $x_0$ and $E_b$ have the same meaning as in Model C.

\begin{table}[h!]
\begin{minipage}[t]{.5\textwidth}
\vspace{3.5pt}
\centering
\begin{dataTable}%
{@{\hspace{2ex}} l @{\hspace{6ex}} l @{\hspace{2ex}}}%
{\scshape\textcolor{white}{Parameter}} &
{\scshape\textcolor{white}{Value}}\\ \midrule[0pt]
$a_\textrm{max}$ 	& 163 bp\\ \midrule
$a_\textrm{min}$ 	& 39 bp\\ \midrule
$E_b$ 				& 13.50 $\textrm{k}_\textrm{B}\textrm{T}$\\ \midrule
$\mu$ 				& -16.13 $\textrm{k}_\textrm{B}\textrm{T}$\\ \midrule
$A$ 				& 2.36 $\textrm{k}_\textrm{B}\textrm{T}$\\ \midrule
$x_0$ 				& 74 bp\\ \midrule
$f$ 				& 0.63
\end{dataTable}
\end{minipage}
\hfill
\begin{minipage}[t]{.45\textwidth}
\vspace{0pt}
\centering
\includegraphics[width=\textwidth]{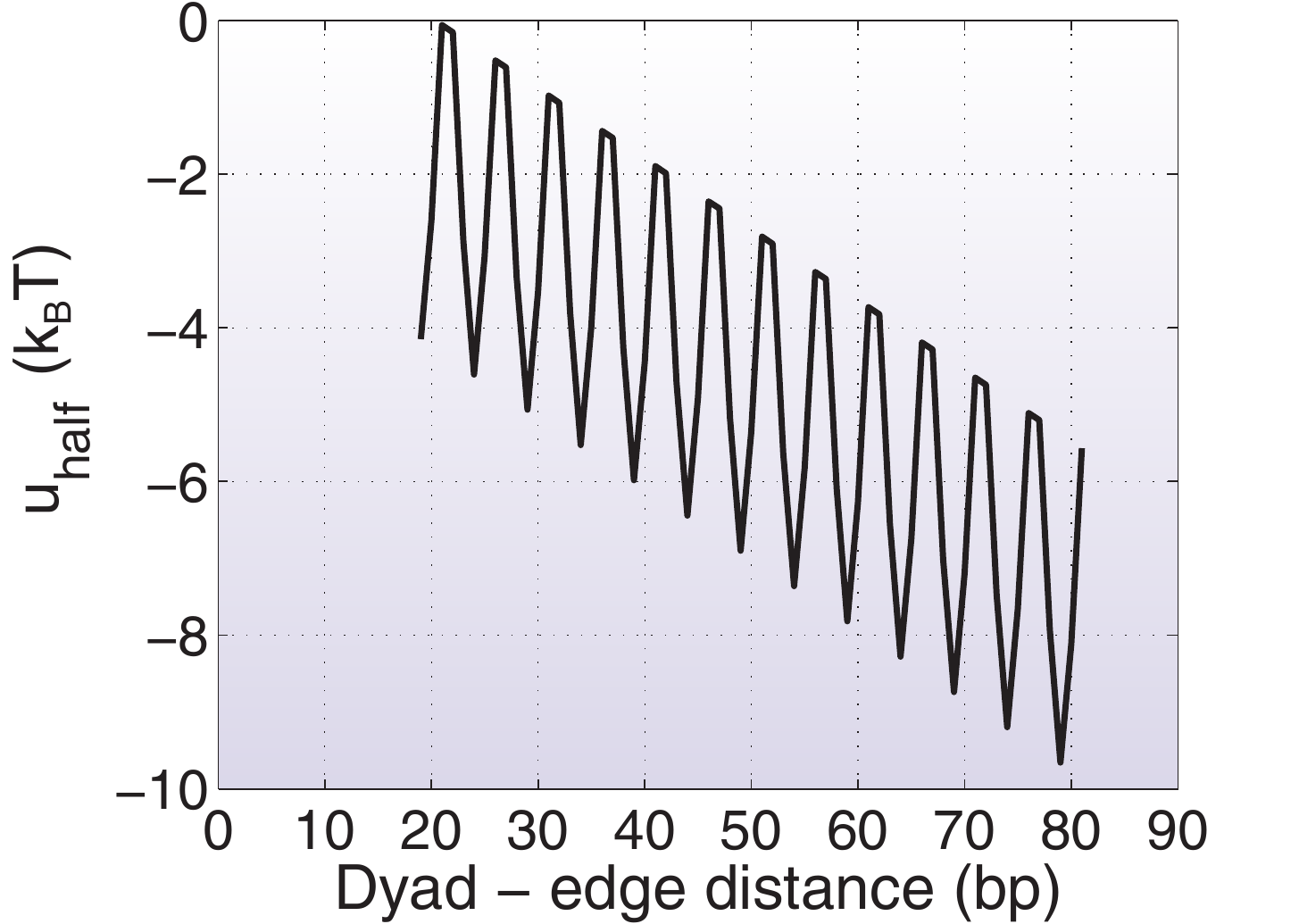}
\end{minipage}
\hspace*{\fill}
\caption{Fitted parameters for Model F. All parameters are as in Model C.}
\end{table}

Fit residuals:
\begin{align*}
&RMS = 9.8984 \times 10^{-4}\\
&r_{\text{osc}} = 0.206240\\
&RMS_{\text{osc}} = 3.0554 \times 10^{-4}
\end{align*}
All residuals are defined as in Model A.

\newpage
{\bf Model G: 5-bp stepwise unwrapping.}
\[
u_\textrm{half}(x) = - E_\textrm{step}\ \texttt{ceil}\left(\frac{x - x_0}{5}\right),
\]
where $E_\textrm{step}$ is the amount of energy lost in each step, and $x_0$ determines the phase of the stepwise profile.

\begin{table}[h!]
\begin{minipage}[t]{.5\textwidth}
\vspace{3.5pt}
\centering
\begin{dataTable}%
{@{\hspace{2ex}} l @{\hspace{6ex}} l @{\hspace{2ex}}}%
{\scshape\textcolor{white}{Parameter}} &
{\scshape\textcolor{white}{Value}}\\ \midrule[0pt]
  $a_\textrm{max}$ & 163 bp\\ \midrule
  $a_\textrm{min}$ & 39 bp\\ \midrule
  $E_\textrm{step}$ & 0.48 $\textrm{k}_\textrm{B}\textrm{T}$\\ \midrule
  $\mu$ & -12.83 $\textrm{k}_\textrm{B}\textrm{T}$\\ \midrule
  $x_0$ & 2 bp\\ \midrule
  $f$ & 0.63
\end{dataTable}
\end{minipage}
\hfill
\begin{minipage}[t]{.45\textwidth}
\vspace{0pt}
\centering
\includegraphics[width=\textwidth]{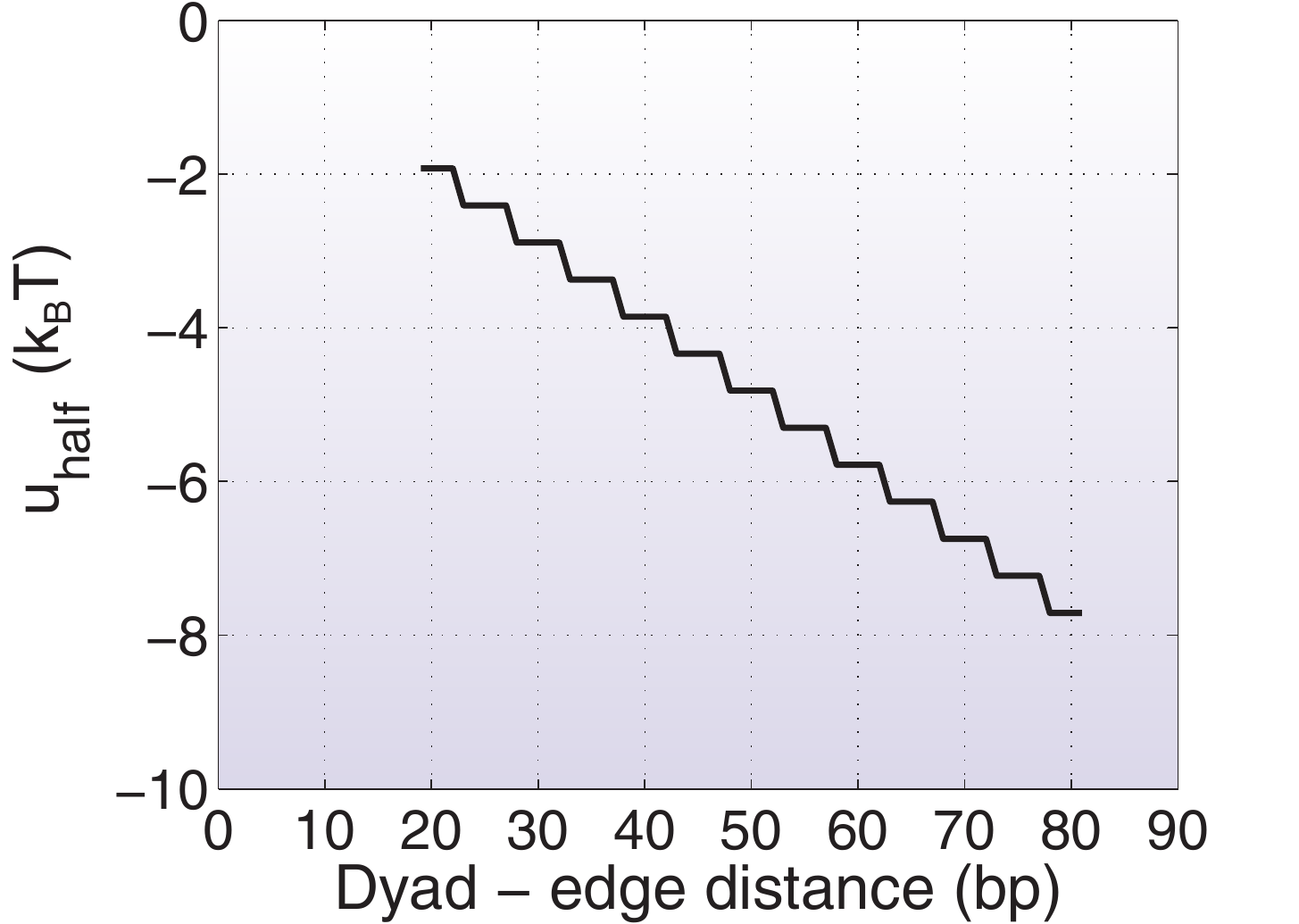}
\end{minipage}
\hspace*{\fill}
\caption{Fitted parameters for Model G. All parameters are as in Model A, except for $E_\textrm{step}$ and $x_0$ defined above.}
\end{table}

Fit residuals:
\begin{align*}
&RMS = 9.9040 \times 10^{-4}\\
&r_{\text{osc}} = 0.282593\\
&RMS_{\text{osc}} = 2.7421 \times 10^{-4}
\end{align*}
All residuals are defined as in Model A.

\newpage
{\bf Model H: 10-bp stepwise unwrapping.}
\[
u_\textrm{half}(x) = - E_\textrm{step}\ \texttt{ceil}\left(\frac{x - x_0}{10}\right),
\]
where $E_\textrm{step}$ is the amount of energy lost in each step, and $x_0$ determines the phase of the stepwise profile.

\begin{table}[h!]
\begin{minipage}[t]{.5\textwidth}
\vspace{3.5pt}
\centering
\begin{dataTable}%
{@{\hspace{2ex}} l @{\hspace{6ex}} l @{\hspace{2ex}}}%
{\scshape\textcolor{white}{Parameter}} &
{\scshape\textcolor{white}{Value}}\\ \midrule[0pt]
  $a_\textrm{max}$ & 169 bp\\ \midrule
  $a_\textrm{min}$ & 3 bp\\ \midrule
  $E_\textrm{step}$ & 1.16 $\textrm{k}_\textrm{B}\textrm{T}$\\ \midrule
  $\mu$ & -12.04 $\textrm{k}_\textrm{B}\textrm{T}$\\ \midrule
  $x_0$ & 3 bp\\ \midrule
  $f$ & 0.62
\end{dataTable}
\end{minipage}
\hfill
\begin{minipage}[t]{.45\textwidth}
\vspace{0pt}
\centering
\includegraphics[width=\textwidth]{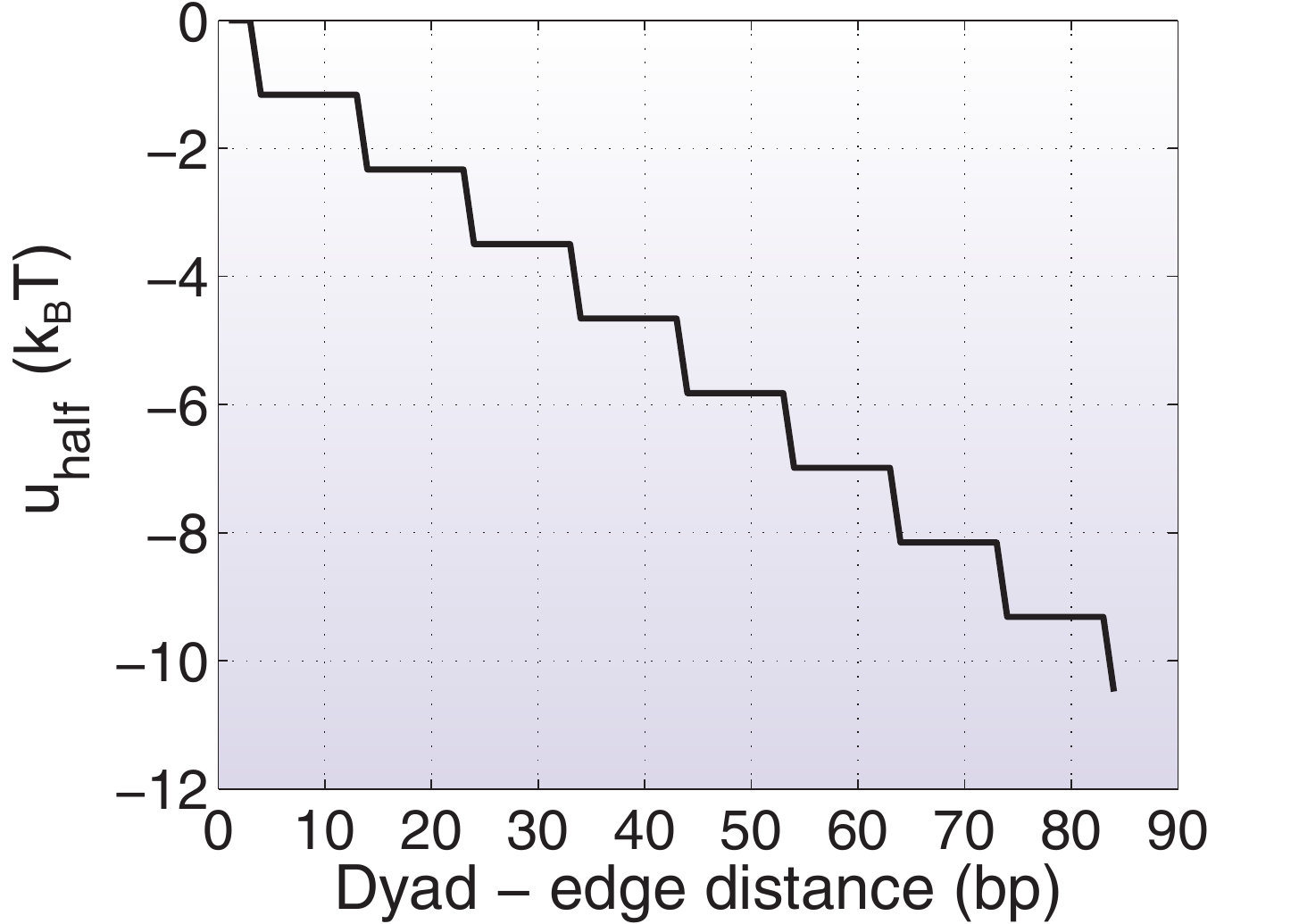}
\end{minipage}
\hspace*{\fill}
\caption{Fitted parameters for Model E. All parameters are as in Model F.}
\end{table}

Fit residuals:
\begin{align*}
&RMS = 9.7895 \times 10^{-4}\\
&r_{\text{osc}} = 0.545077\\
&RMS_{\text{osc}} = 2.4569 \times 10^{-4}
\end{align*}
All residuals are defined as in Model A.

\newpage
\section{Supplementary Figures}
\begin{figure}[ht!]
\centerline{
\includegraphics[width=5in]{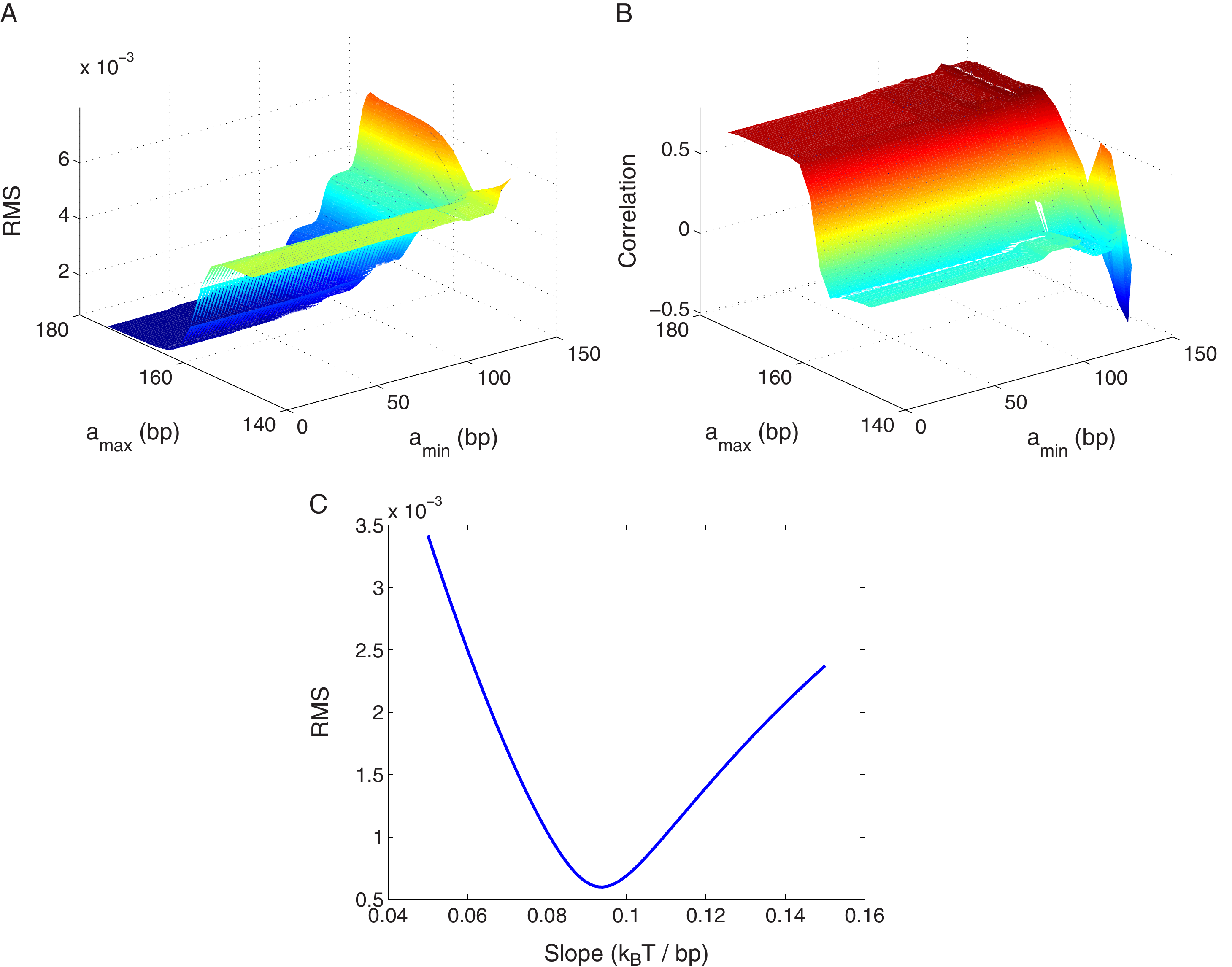}
}
\caption{\textbf{Sensitivity of the predicted inter-dyad distribution to the parameters of the unwrapping potential based on nucleosome crystal structures.}
(A) Root-mean-square error (RMS) of the inter-dyad distribution predicted using the model in Fig.~1B, as a function of $a_\textrm{min}$ and $a_\textrm{max}$.
(B) The linear correlation coefficient between oscillations in the predicted and observed inter-dyad distributions ($r_\text{osc}$), as a function of $a_\textrm{min}$ and $a_\textrm{max}$.
The oscillations  were obtained by subtracting the smooth background from inter-dyad distributions, as described in the Fig.~1 caption.
(C) Variation of the RMS with the slope of the unwrapping potential in Fig.~1B.
In all panels, model parameters that were not varied were kept fixed at their best-fit values (SI Results, Model A).
}
\end{figure}

\begin{figure}[ht!]
\centering
\includegraphics[width=3in]{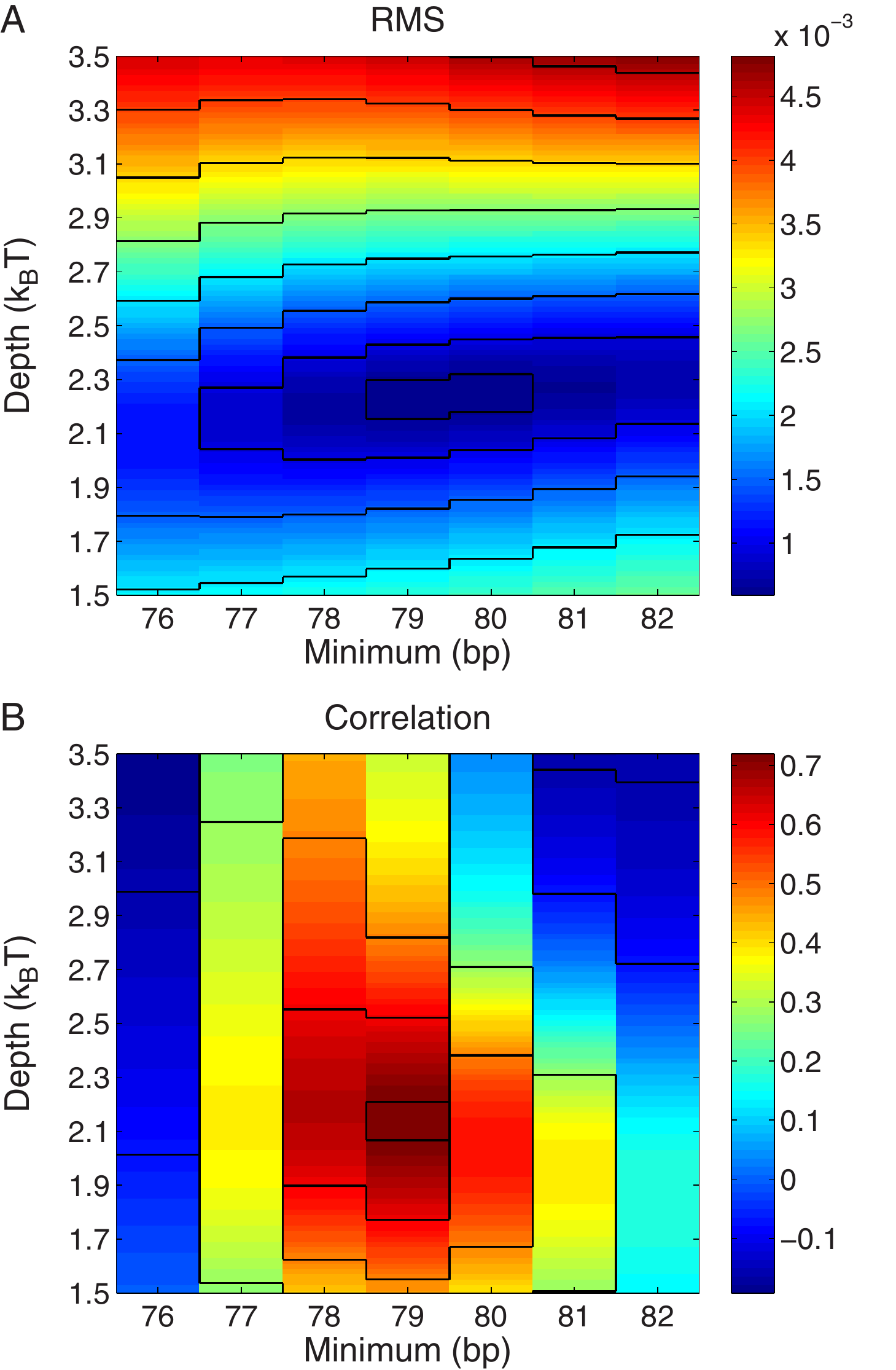}
\caption{\textbf{Sensitivity of the predicted inter-dyad distribution to model parameters describing higher-order chromatin structure.}
The unwrapping potential is based on nucleosome crystal structures (SI Results,
Model A).
(A) Root-mean-square error (RMS) of the predicted inter-dyad distribution, as a function of the position and the depth of the first minimum
outside of the nucleosome core (Fig.~1B). The depth of the first minimum is computed with respect to $u_\textrm{half}(x = 73 \text{ bp})$.
(B)  The linear correlation coefficient between oscillations in the predicted and observed inter-dyad distributions ($r_\text{osc}$), as a function of
the position and the depth of the first minimum outside of the nucleosome core (Fig.~1B).
The oscillations  were obtained by subtracting the smooth background from inter-dyad distributions, as described in the Fig.~1 caption.
In both panels, model parameters that were not varied were kept fixed at their best-fit values (SI Results, Model A).
}
\end{figure}

\begin{figure}[ht!]
\centerline{
\includegraphics[width=5.5in]{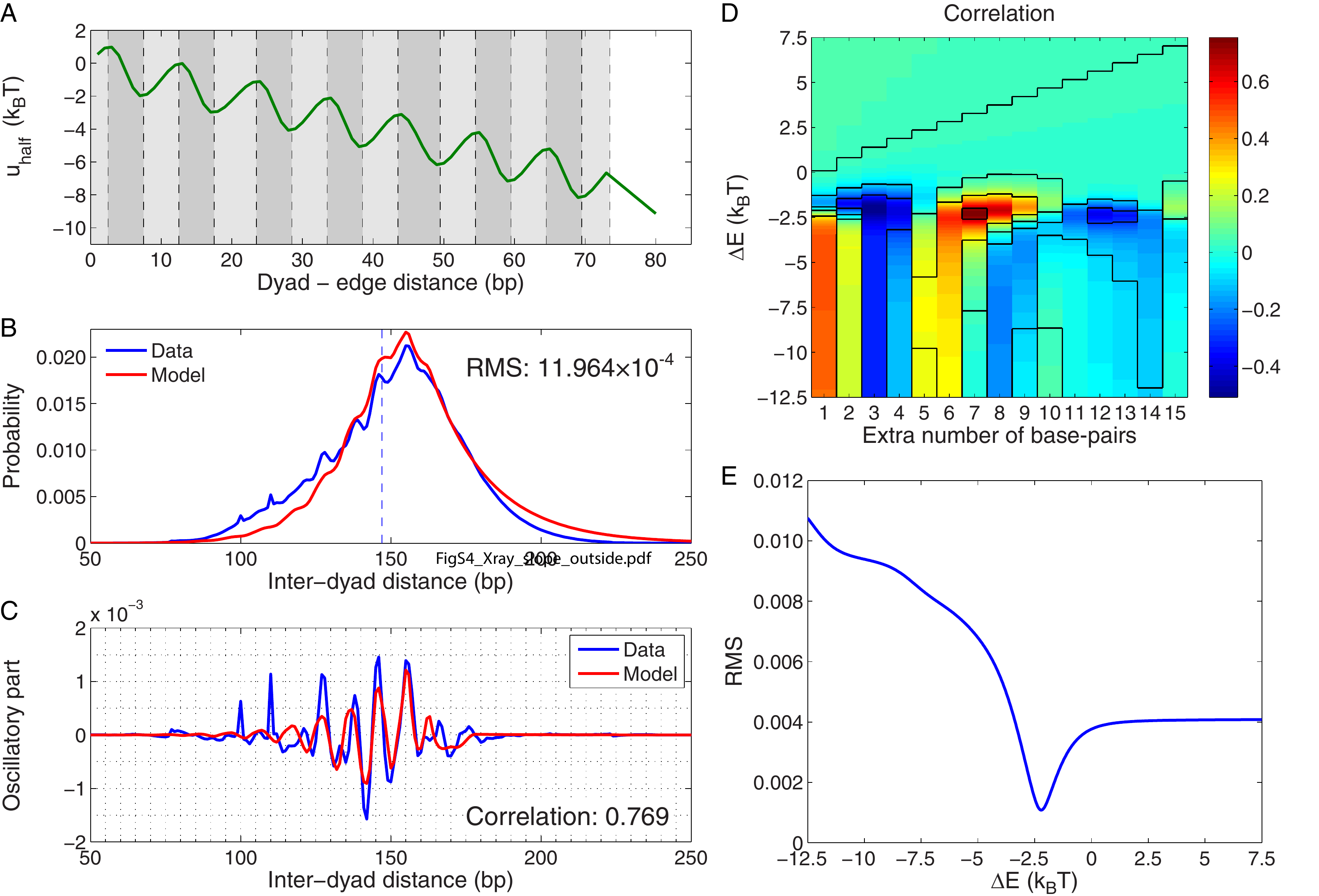}
}
\caption{\textbf{Crystal structure-based model augmented by a linear potential outside of the nucleosome core.}
(A) The energy profile fitted to reproduce the inter-dyad distance distribution shown in (B). All fitting parameters are listed in SI Results, Model B.
Under the symmetric unwrapping assumption, the energy of a nucleosome which covers $2x + 1$~bps is given by $2 u_\textrm{half}(x)$.
(B) The inter-dyad distance distribution observed in a high-resolution nucleosome map~\cite{Brogaard2012} (blue line), and predicted using Model B in SI Results (red line).
RMS - root-mean-square deviation between the model and the data. 
Note that in this model RMS below $10^{-3}$ could not be achieved, and thus optimization was switched to maximize the correlation coefficient $r_\text{osc}$
once RMS reached $1.2 \times 10^{-3}$ (see SI Methods for details).
(C) Oscillations in the observed (blue line) and predicted (red line) inter-dyad distributions. The oscillations were obtained
by subtracting the smooth background from the data and the model in (B), as described in the Fig.~1 caption.
Correlation refers to $r_\text{osc}$, the linear correlation coefficient between measured and predicted oscillations.
(D) Heatmap with superimposed contour lines of the $r_\text{osc}$ dependence on the two parameters of the linear potential
outside of the nucleosome core: $\Delta x = x_{\text{last}} - 73$~bp and $\Delta E = u_\textrm{half}(x_{\text{last}}) - u_\textrm{half}(73)$,
where $[1, x_{\text{last}}]$ is the range of the energy profile (SI Results, Model B). Note that the best fit corresponds to $\Delta x = 7$~bp.
(E) The dependence of the RMS on $\Delta E$ for the best-fit value of $\Delta x = 7$~bp.
All parameters not explicitly varied in (D) and (E) were kept fixed at their best-fit values (SI Results, Model B).
}
\end{figure}

\begin{figure}[ht!]
\centerline{
\includegraphics[width=5.5in]{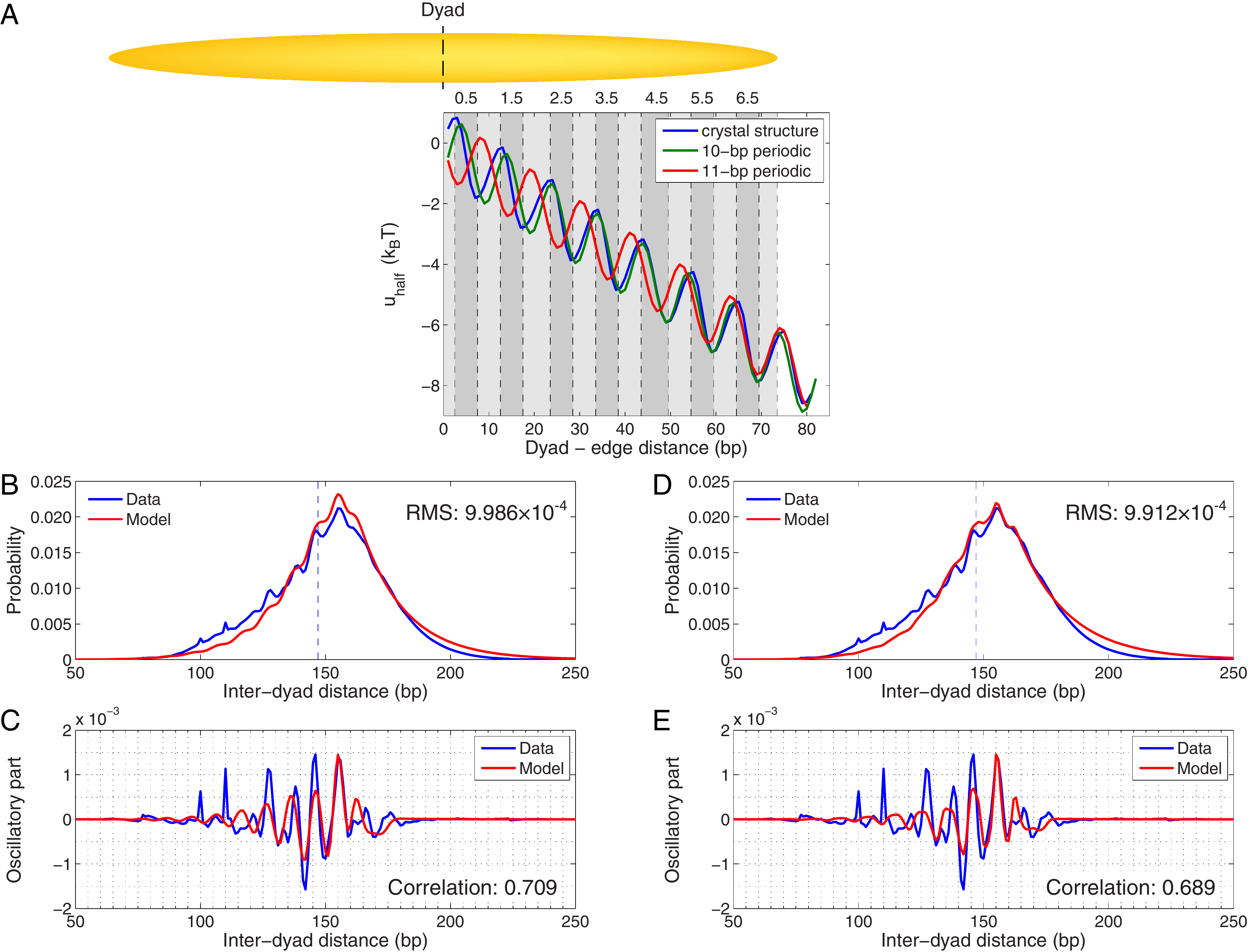}
}
\caption{\textbf{Strictly periodic models of nucleosome unwrapping.}
(A) Nucleosome unwrapping/higher-order structure potential energy profiles.
Under the symmetric unwrapping assumption, the energy of a nucleosome that covers $2x + 1$~bps is given by $2 u_\textrm{half}(x)$.
The minima and maxima of the energy landscape are either based on the crystal structures of the nucleosome core particle as in Figure~1 (blue), or else are 10 (green) and 11 (red) bp-periodic
oscillations with fitted initial phase (SI Results, Models C and D).
Dark gray bars show where the histone binding motifs interact with the DNA minor groove.
Light gray bars indicate where the DNA major groove faces the histones.
(B) The inter-dyad distance distribution from a high-resolution nucleosome map ~\cite{Brogaard2012} (blue line),
and from the 10 bp-periodic model (red line). All model parameters are listed in SI Results, Model C.
RMS - root-mean-square deviation between the model and the data. 
(C) Oscillations in the observed (blue line) and predicted (red line) inter-dyad distributions. The oscillations were obtained by subtracting a smooth background from the data and the model in (B), as described in the Fig.~1 caption.
Correlation refers to $r_\text{osc}$, the linear correlation coefficient between measured and predicted oscillations.
(D) Same as (B), for the 11 bp-periodic model. All model parameters are listed in SI Results, Model D.
(E) Same as (C), for the 11 bp-periodic model.
}
\end{figure}

\begin{figure}[ht!]
\centering
\includegraphics[width=3in]{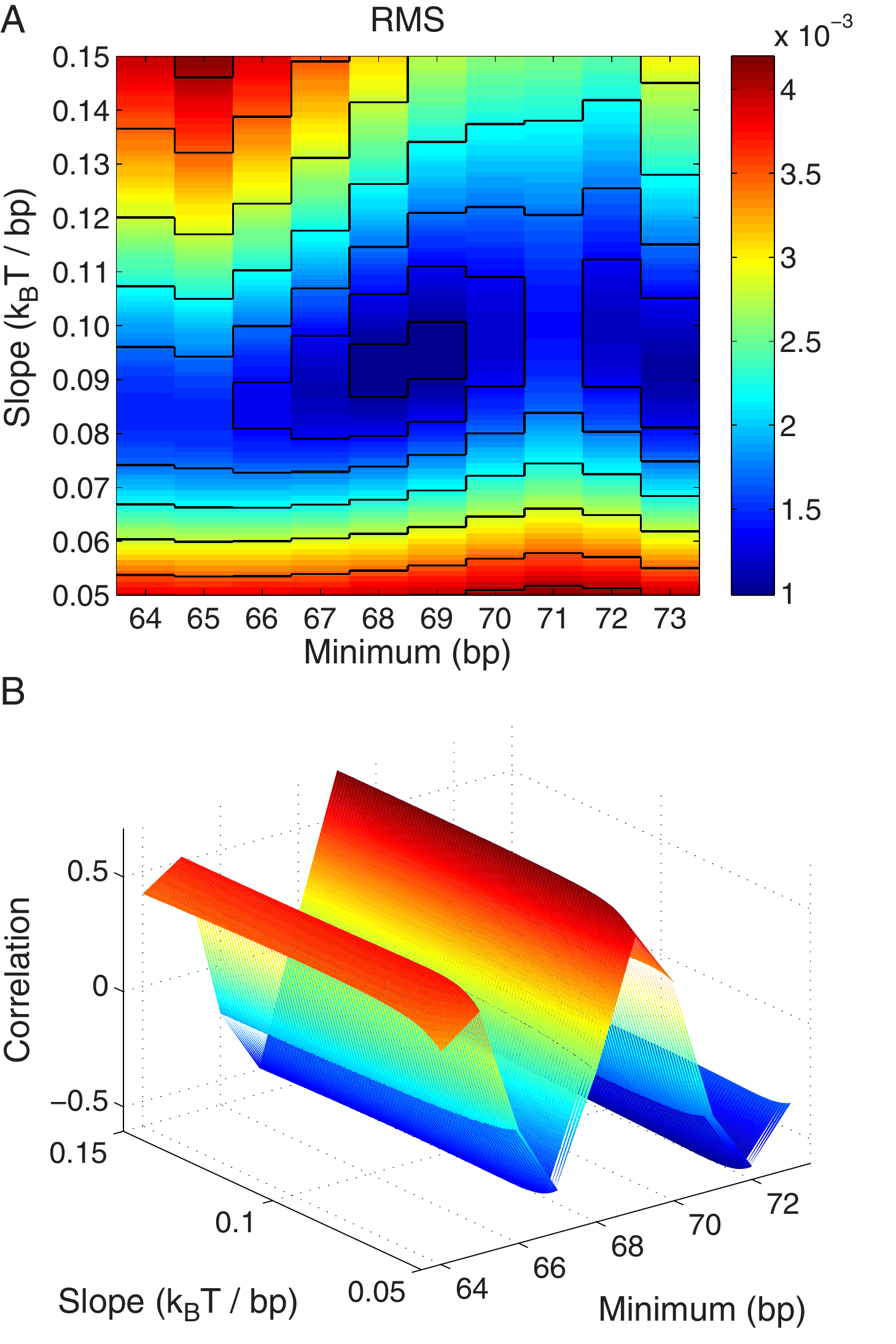}
\caption{\textbf{Sensitivity of the predicted inter-dyad distribution to parameters of the 10 bp-periodic model.}
(A) Heatmap with superimposed contour lines of the RMS dependence on the slope of the energy profile and the position of the last minimum within the nucleosome core.
RMS - root-mean-square deviation between the model and the data. 
(B) The linear correlation coefficient $r_\text{osc}$ between oscillations in the predicted and observed inter-dyad distributions, as a function of
the overall slope of the energy profile and the position of the last minimum within the nucleosome core.
All parameters not explicitly varied were kept fixed at their best-fit values (SI Results, Model C).
}
\end{figure}

\begin{figure}[ht!]
\centerline{
\includegraphics[width=8.2in]{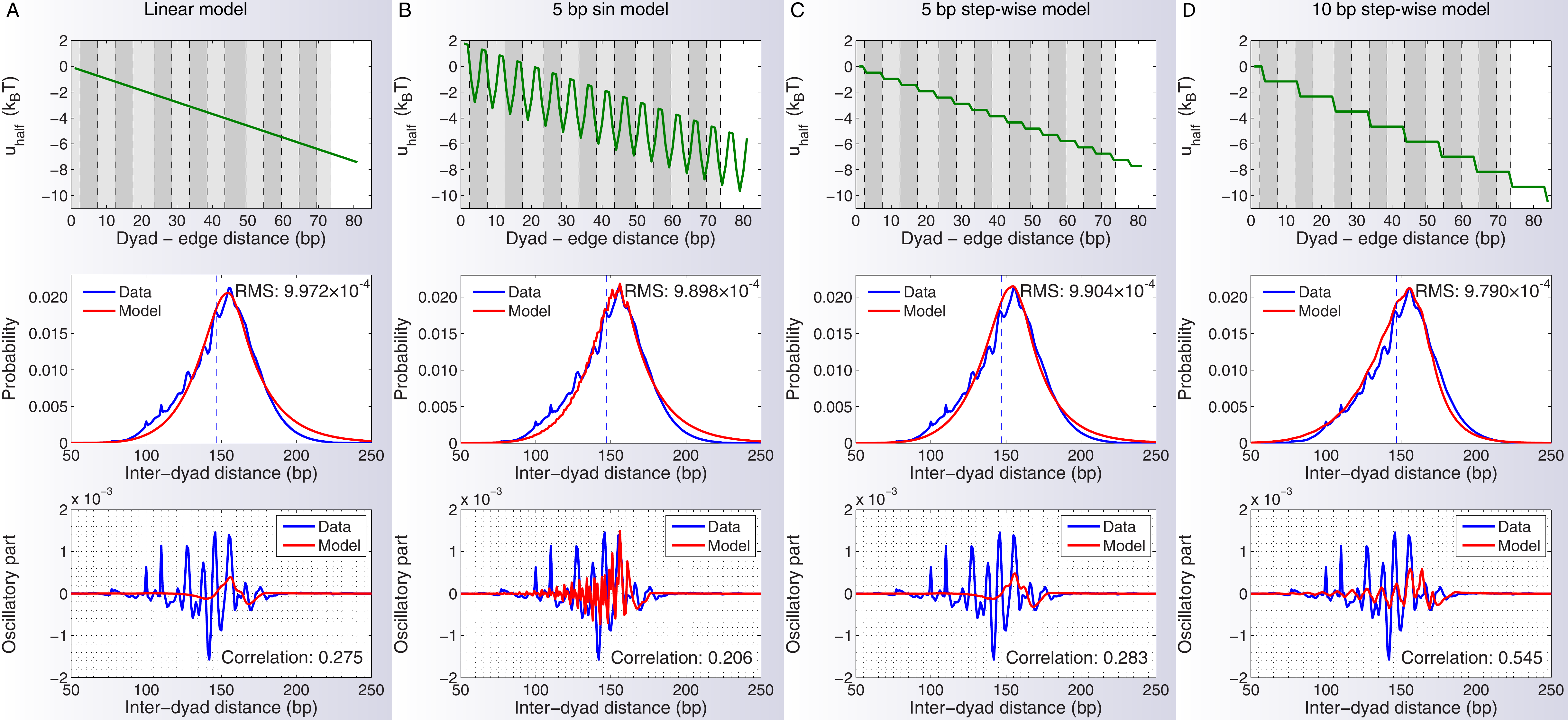}
}
\caption{\textbf{Alternative models of nucleosome unwrapping.}
(A) Linear model (SI Results, Model E). (B) 5-bp periodic model (SI Results, Model F).
(C) 5-bp step-wise model (SI Results, Model G). (D) 10-bp step-wise model (SI Results, Model H).
In each column, the upper panel shows the nucleosome unwrapping/higher-order structure potential energy profile (as in Fig.~1B), the middle panel
shows the comparison of experimental and predicted inter-dyad distance distributions (as in Fig.~1C), and the lower panel shows observed and predicted oscillations in the inter-dyad distance distributions (as in Fig.~1D).
}
\end{figure}

\begin{figure}[ht!]
\centerline{
\includegraphics[width=6.0in]{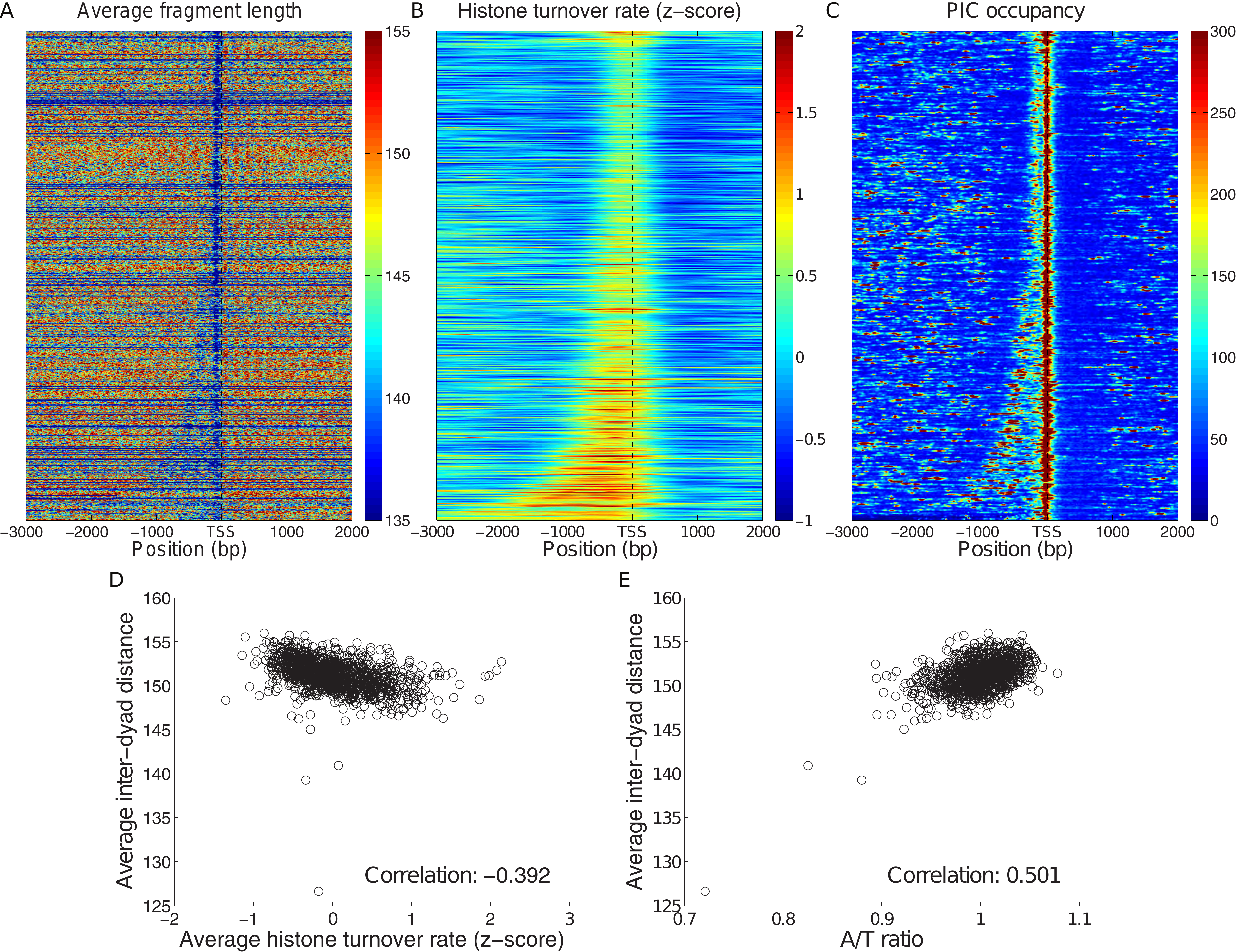}
}
\caption{\textbf{Genome-wide distribution of nucleosome lengths, histone turnover rates, and transcription pre-initiation complexes.}
(A) Distribution of average lengths of DNA-bound particles mapped by MNase digestion~\cite{henikoff:2011} in the vicinity of TSS. We considered particles with sizes between 80 and 200 bp and assigned particle lengths to the mid-point of each particle.
Values for bps without dyads were obtained by interpolation.
(B) Distribution of histone turnover rates~\cite{dion:2007} in the vicinity of TSS.
(C) Distribution of the combined occupancy of 9 transcription pre-initiation complexes (PICs)~\cite{rhee:2012} in the vicinity of TSS. PIC occupancies provided at 20 bp interval in Ref.~\cite{rhee:2012} were interpolated.
In panels (A)-(C), gene order is as in Figure~1B, and the heatmaps were smoothed using a 2D Gaussian kernel with $\sigma = 3$ pixels.
(D) Correlation between inter-dyad distances and histone turnover rates averaged
over $10$ kbp windows tiling the yeast genome.
(E) Correlation between average inter-dyad distances and the A/T ratio
in $10$ kbp windows tiling the yeast genome. A/T ratio is the fraction of A/T nucleotides in the window, divided by the genome-wide A/T fraction. Correlation in (D) and (E) refers to the
linear correlation coefficient.
}
\end{figure}

\begin{figure}[ht!]
\centerline{
\includegraphics[width=4.3in]{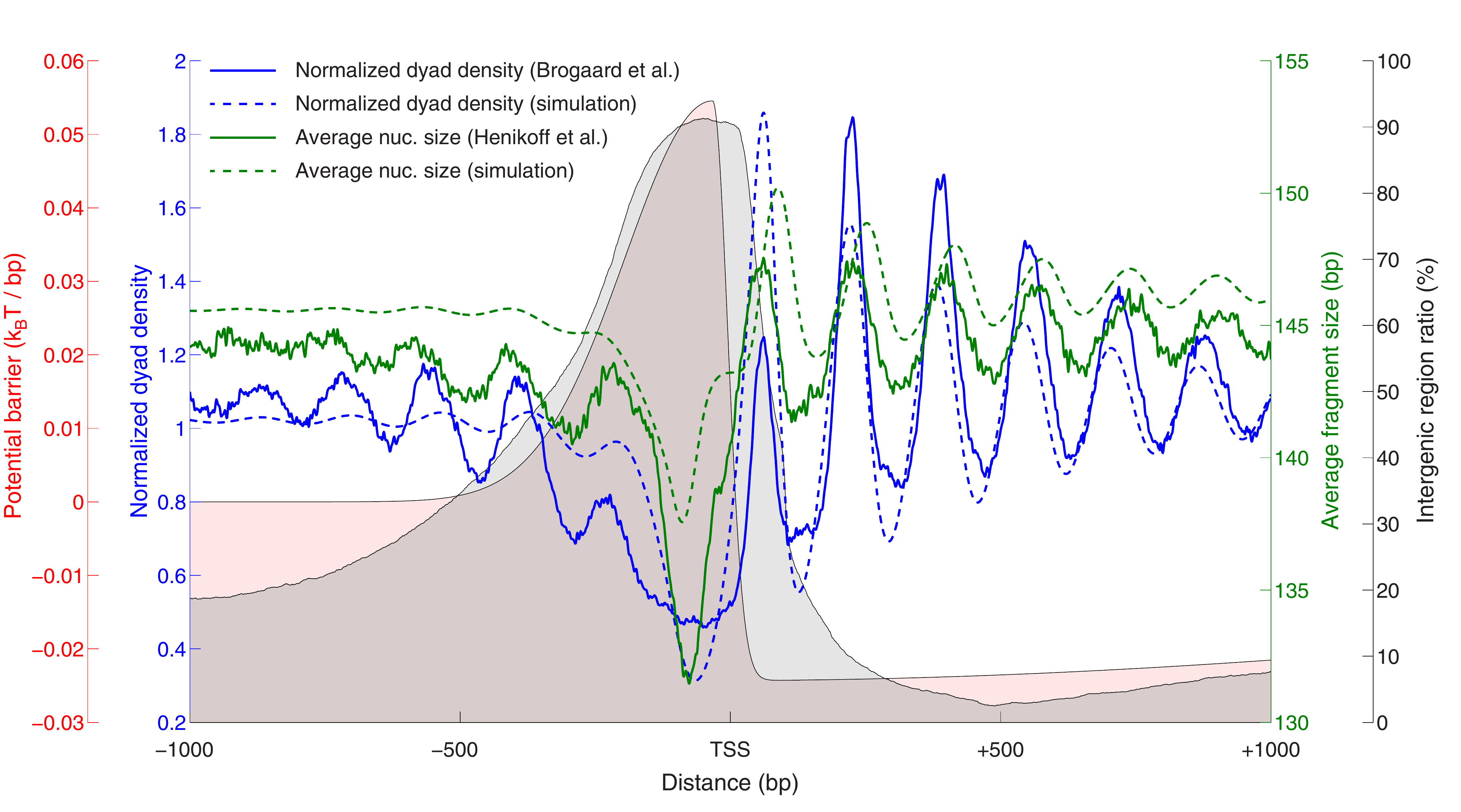}
}
\caption{\textbf{Modeling distributions of nucleosome lengths and dyad positions in the vicinity of TSS.} We align all yeast genes by their TSS as in Fig.~2 and for each bp compute the fraction of times that bp is found in an intergenic region, as opposed to the ORF of a neighboring gene (grey background curve). The intergenic fraction has an asymmetric shape with a maximum at about 50 bp upstream of the TSS. We use this shape as a guide for constructing an energy barrier for \textit{in vivo} histone deposition (pink background curve).
The barrier is composed of three half-Gaussians:
$B(x) = H \exp \left[-\frac{(x-c)^2}{2 \sigma_1^2}\right]~(x \leq c),
~(H + D) \left\{ \exp \left[-\frac{(x-c)^2}{2 \sigma_2^2}\right] - 1\right\} -
D \exp \left[-\frac{(x-c)^2}{2 \sigma_3^2}\right]~(x > c)$.
The free parameters of the barrier are fit to maximize the sum of two correlations:
between observed and predicted normalized dyad counts~\cite{Brogaard2012} (solid and dashed blue lines, respectively), and between observed and predicted average nucleosome DNA lengths~\cite{henikoff:2011} (solid and dashed green lines, respectively).
Normalized dyad counts are computed as the total number of dyads at a given bp for all genes,
divided by the average of this quantity in a [-1000,1000] bp window around the TSS.
Average DNA lengths are computed for all nucleosomes with a midpoint at a given bp, for all
genes (if the midpoint falls in between two bps, the one on the left is used).
The fitted parameters are: $H = 0.0545\text{ k}_\text{B}\text{T}, D = 0.0243\text{ k}_\text{B}\text{T},
c = x_\text{TSS} - 32 \text{ bp}, \sigma_1 = 162.7 \text{ bp},
\sigma_2 = 28.0 \text{ bp}, \sigma_3 = 2090.9 \text{ bp}$,
where $x_\text{TSS}$ is the absolute position of the TSS in the box, $c$ is the center of the 3 Gaussians, $H$ is the height of the first Gaussian, $D$ is the depth of the third Gaussian, and $\sigma_1, \sigma_2$, $\sigma_3$ are the standard deviations of the three Gaussian distributions. The simulations were done in a 15 kbp box with the barrier placed at its center to eliminate the boundary effects. Unwrapping was assumed to be symmetric and the nucleosome structure-based unwrapping potential (SI Results, Model A) was used. The total free energy $u_\text{nuc} (k,l)$ of a nucleosome occupying bps $k, \ldots ,l$ is a sum
of $u^\text{SI}_\text{nuc}$ and $u^\text{barrier}_\text{nuc} = \sum_{j=k}^{l} \epsilon_j$, where $\epsilon_j$
is the value of the barrier at bp $j$.
Note that the chemical mapping data underestimates the number of -1 and +1 nucleosomes due to gel selection bias~\cite{Brogaard2012}.
}
\end{figure}

\begin{figure}[ht!]
\centering
\includegraphics[width=3in]{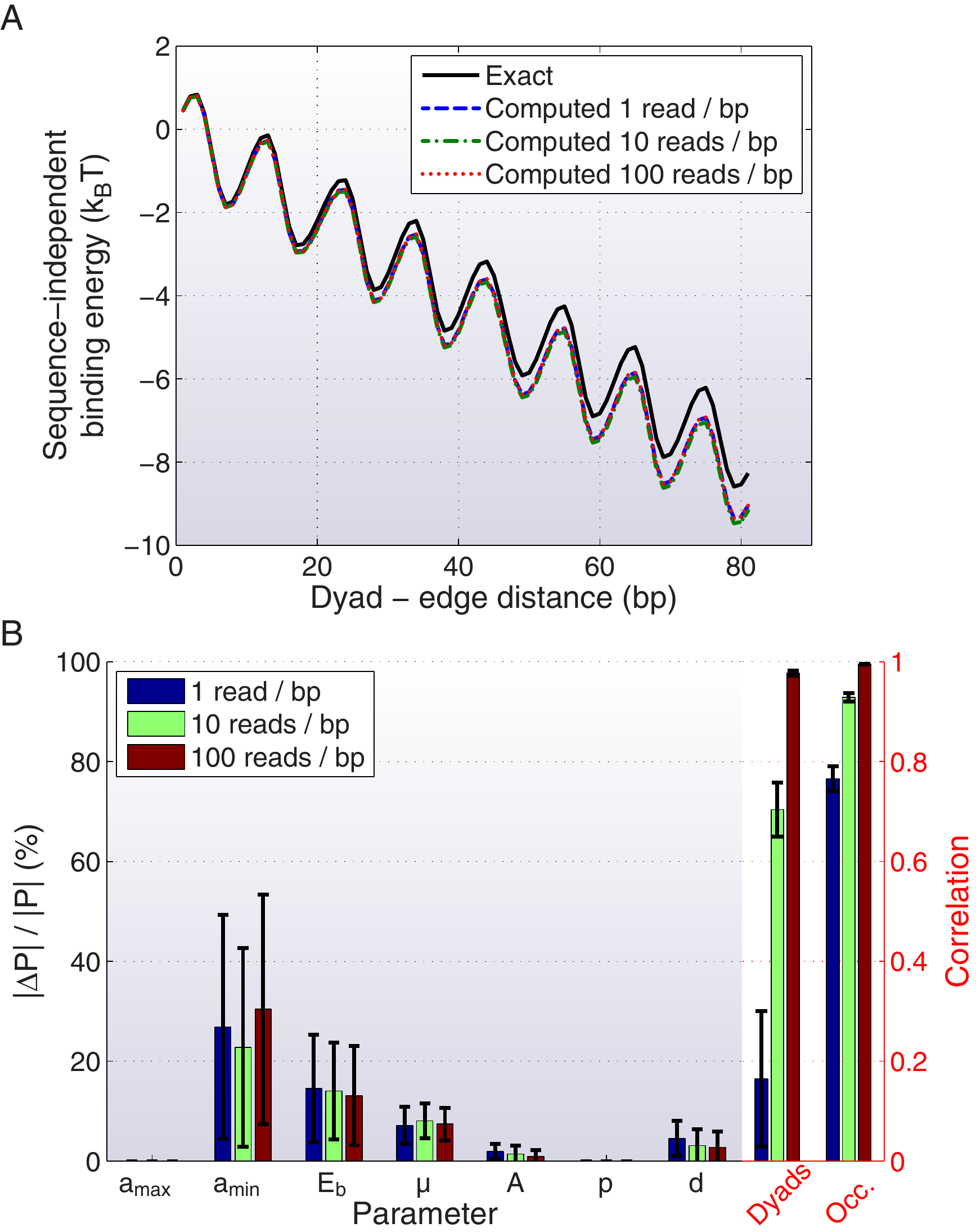}
\caption{\textbf{Inference of the unwrapping potential in a model with sequence-specific
nucleosome formation energies.}
(A) Exact unwrapping potential vs. unwrapping potentials predicted at three levels of
sequence coverage. All calculations were done on \textit{S. cerevisiae} chromosome I with
$L = 230$ kbp. $M L$ reads were randomly sampled from the exact distribution
$n_1^\text{nuc} (i,j)$ (see Materials and Methods), where $M \in \{1,10,100\}$ is the desired
level of read coverage per bp.
Sampled reads were used to compile a chromosome-wide histogram of nucleosome DNA lengths,
to which the unwrapping potential in SI Results, Model A was fit using a
genetic algorithm optimization function \texttt{ga} from MATLAB, which minimizes the root-mean-square error of the predicted distribution of nucleosome lengths.
(B) The relative errors between predicted and exact (SI Results, Model A) parameters of the unwrapping potential at three levels of sequence coverage (light blue background). $P$ denotes any parameter on the horizontal axis. Linear correlation between predicted and exact
distributions of dyad positions and nucleosome occupancy (light pink background).
The height of each bar represents the mean relative error for the corresponding parameter or the mean correlation coefficient, obtained by averaging the results of 100 random sampling experiments. The uncertainty intervals represent standard deviations.
}
\end{figure}
\clearpage
\bibliographystyle{ieeetr}
\bibliography{Unwrapping_main,Unwrapping_SI}

\end{document}